\documentclass[twocolumn]{aastex701}
\usepackage{graphicx}
\usepackage{mathtools}
\usepackage{lineno}
\usepackage{multirow}
\usepackage{siunitx}
\usepackage{adjustbox}
\usepackage{booktabs} 
\usepackage{amsmath}
\usepackage{amssymb}
\usepackage{tikz}
\usetikzlibrary{angles,quotes,decorations.markings,arrows.meta,calc}
\usetikzlibrary{decorations.markings}
\linenumbers
\received{July 08, 2026}
\revised{August 18, 2026}
\accepted{August 22, 2026}

\shorttitle{JWST FGK Limb Darkening Survey}
\shortauthors{Sing et al.}
\graphicspath{{./}{figures/}}

\begin{document}

\title{A JWST transiting survey of FGK stellar limb darkening: empirical evidence for quadratic laws and atmospheric model comparisons}


\correspondingauthor{D. K. Sing}
\email{dsing@jhu.edu}

\author[0000-0001-6050-7645]{David K. Sing}
\affiliation{Department of Earth \& Planetary Sciences, Johns Hopkins University, Baltimore, MD, USA}
\affiliation{William H.\ Miller III Department of Physics \& Astronomy, Johns Hopkins University, 3400 N Charles St, Baltimore, MD, USA}
\email{dsing@jhu.edu}

\author[0000-0003-3667-8633]{Joshua D. Lothringer}
\affiliation{Space Telescope Science Institute, Baltimore, MD 21218, USA}
\email{jlothringer@stsci.edu}

\author[0000-0003-3305-6281]{Jeff A. Valenti}
\affiliation{Space Telescope Science Institute, Baltimore, MD 21218, USA}
\email{valenti@stsci.edu}


\author[0000-0002-0832-710X]{Natalie H. Allen}\email[show]{nallen19@jhu.edu}
\affiliation{William H.\ Miller III Department of Physics \& Astronomy, Johns Hopkins University, 3400 N Charles St, Baltimore, MD, USA}

\author[orcid=0000-0002-9030-0132]{Katherine A. Bennett}
\affiliation{Department of Earth \& Planetary Sciences, Johns Hopkins University, Baltimore, MD, USA}
\email[show]{kbenne50@jhu.edu}  

\author[0000-0001-5097-9251]{Carlos Gascón}
\affiliation{Space Telescope Science Institute, Baltimore, MD 21218, USA}
\email{cgascon@stsci.edu}

\author[0000-0002-1624-3360]{Mei Ting Mak}
\altaffiliation{Croucher Postdoctoral Fellow}
\email{martha.mak@physics.ox.ac.uk}
\affiliation{Atmospheric, Oceanic, and Planetary Physics Department, University of Oxford, OX1 3PU, UK}
\affiliation{Department of Physics and Astronomy, Faculty of Environment, Science and Economy, University of Exeter, Exeter EX4 4QL, UK}

\author[0000-0003-0473-6931]{Patrick McCreery}
\affiliation{William H.\ Miller III Department of Physics \& Astronomy, Johns Hopkins University, 3400 N Charles St, Baltimore, MD, USA}
\email{pmccree2@jhu.edu}

\author[0000-0003-1622-1302]{Sagnick Mukherjee}\email[show]{smukhe50@asu.edu}
\altaffiliation{51 Pegasi b Fellow} 
\affiliation{School of Earth and Space Exploration, Arizona State University, Tempe, AZ, USA \\ }

\author[0000-0002-1056-3144]{Lakeisha M. Ramos Rosado}
\affiliation{William H.\ Miller III Department of Physics \& Astronomy, Johns Hopkins University, 3400 N Charles St, Baltimore, MD, USA}
\email{lramosr1@jhu.edu}

\author[0000-0001-8510-7365]{Stephen P.\ Schmidt}
\altaffiliation{NSF Graduate Research Fellow}
\affiliation{William H.\ Miller III Department of Physics \& Astronomy, Johns Hopkins University, 3400 N Charles St, Baltimore, MD, USA}
\email{sschmi42@jhu.edu}

\author[0000-0002-7352-7941]{Kevin B. Stevenson}
\affiliation{Johns Hopkins APL, 11100 Johns Hopkins Rd, Laurel, MD 20723, USA}
\email{Kevin.Stevenson@jhuapl.edu}

\author[0000-0002-5113-8558]{Daniel P. Thorngren}
\affiliation{William H. Miller III Department of Physics and Astronomy, Johns Hopkins University, Baltimore, MD 21218, USA}
\email{dpthorngren@jhu.edu}

\author[0000-0002-6379-3816]{Le-Chris Wang}
\affiliation{Department of Astrophysical Sciences, Princeton University, 4 Ivy Lane, Princeton, NJ 08544, USA}
\email{lechris.wang@princeton.edu}


\begin{abstract}
\noindent We present a study of stellar limb-darkening using JWST transit observations of seven exoplanets orbiting FGK host stars, spanning 4200-6800 K. The wide wavelength coverage and high S/N of NIRISS/SOSS and NIRSpec/PRISM enable precise constraints on the wavelength-dependent limb-darkening. Using Bayesian model selection, 
we find that the quadratic limb-darkening law is statistically preferred over higher-order laws for most FGK stars, recovers consistent intensity profiles for $>$95\% of the stellar disk area, and introduces a minimal bias in the derived transit depths of only $\sim$14 ppm (1$\sigma$) for 6/7 targets. This contrasts with previous studies relying on stellar models.
We compare the empirically derived quadratic coefficients to predictions from the PHOENIX, MPS-ATLAS, MURaM, and Stagger stellar atmosphere grids. 
We introduce a quadratic limb-darkening parameterization in terms of limb intensity ($\ell$) and curvature at mid-$\mu$ ($\delta$), finding that empirical FGK limb darkening is generally more linear than models predict ($\delta \lesssim 0.1$).
We identify wavelength-independent offsets between data and model quadratic coefficients, minimized by adopting $\mu_{\rm min} = 0.2$ in intensity calculations; we attribute this in part to models overpredicting limb-darkening near the limb where the plane-parallel approximation breaks down.
For spherical PHOENIX models, we derive a $\mu$ rescaling method using a $\tau = 1$ photospheric radius. With these corrections, residual offsets are minimized and all stellar models achieve statistically acceptable fits.
We provide recommended limb-darkening offset priors for use in JWST transit analyses, enabling more accurate constraints on exoplanet transmission spectra while accounting for residual stellar model uncertainties.
 \end{abstract}




\section{Introduction} \label{sec:intro}

For more than a quarter of a century, transit events have been used to revolutionize our understanding of exoplanets and their host stars \citep{Charbonneau2000,2000ApJ...529L..41H}.  Exquisite transit photometry from space telescopes such as the Hubble Space Telescope (HST), Kepler, NASA's Transiting Exoplanet Survey Satellite (TESS), and now the James Webb Space Telescope (JWST) have unlocked a wealth of astrophysical information, including planet radii, atmospheric composition, and temperature structure as well as fundamental host star parameters like the stellar mean density \citep{2003ApJ...585.1038S, 2017JGRE..122...53D}.
All of this vital information depends on how well the stellar limb darkening can be modeled, as it has a strong wavelength-dependent effect on the shape of the transit light curve \citep{Brown2001,MandelAgol2002,Knutson2007}.

An exoplanet host star's atmospheric temperature structure and opacity impact its planets' transit light curves through stellar limb darkening (LD).
Closer to the stellar limb, the line-of-sight path within the atmosphere increases relative to the geometry of the center of the stellar disk; this raises the height at which the atmosphere becomes opaque and therefore the photosphere.
As a stellar atmosphere is cooler at higher altitudes, cooler and dimmer stellar layers are seen at the limbs, making them relatively darker than at the center of the stellar disk. 
For a transit light curve, limb darkening creates a characteristic U-shape with a `rounded' bottom at mid-transit as the planet is blocking relatively less light closer to the limb. 
The effects are strongest at shorter wavelengths as the difference in blackbody temperatures and resulting emission is strongest. 
At longer infrared wavelengths in the Rayleigh-Jeans tail, the difference in emission between stellar atmospheric layers radiating at different temperatures becomes negligible; the resulting limb-darkening is weak and transit light curves have a more box-like shape. With JWST's wide wavelength coverage, these effects can be seen within a single transit \citep[e.g., WASP-39~b, NIRSpec/PRISM, 0.5 to 5.5 $\mu$m;][]{Rustamkulov2023Natur.614..659R}.

Different functional forms describing the wavelength-dependent change in intensity of the star's radiation across the disk, usually referred to as limb-darkening laws for historic reasons, have been used when modeling transit light curves \citep[e.g.,][]{MandelAgol2002}.
Popular functional forms include linear law, quadratic, 3-parameter nonlinear \cite{Sing2009},  4-parameter nonlinear law \citep{Claret2000} and the power-2 law \citep{2017AJ....154..111M}. 
The parameters for these laws can either be freely fit in a transit light curve or constrained to values calculated using atmospheric models \citep[e.g.,][]{2010A&A...510A..21S,2015MNRAS.453.3821P,2024JOSS....9.6816G}. 
Quadratic laws are the most commonly employed to fit transit data, with previous studies indicating wavelength-dependent offsets or biases in the transit depth may occur when adopting this LD law \citep{2024ApJ...977L...7K}. However, these studies have been largely theoretical and rely on the assumption that the atmospheric models themselves accurately reproduce stellar LD. 
If stellar atmosphere models systematically overpredict the limb-darkening curvature, then the biases predicted by such studies may not apply to real observations.
The stellar models themselves require higher-order functional forms to reproduce the model intensities, especially near the limb where the intensity can change rapidly \citep{Claret2000}. However, since the first high-quality HST transit light curves of HD~209458~b became available \citep{2001ApJ...552..699B, Charbonneau2002}, it has been clear that stellar atmospheric models themselves can often struggle to adequately model the wavelength dependent intensity profile of the star \citep{Knutson2007,2008ApJ...686..658S}. 
Improvements have been made over 1D stellar atmospheric models with 3D hydrodynamic models \citep{2012A&A...539A.102H,2015A&A...573A..90M}, though the model parameter space available is limited.

Both the limb darkening laws and the atmospheric models can be empirically tested when compared to high precision limb darkening light curves \citep{2011MNRAS.417.2166S}. Unfortunately, full phase coverage of a transit light curve is typically not measured from the HST, as it is in low-Earth orbit, and studies using transit photometry from CoRoT, Kepler or TESS can sample hundreds of stars but do not test the wavelength dependence of limb darkening \citep[e.g.,][]{2010A&A...510A..21S,2015MNRAS.450.1879E,2023MNRAS.519.3723M}, which is critical to understand when interpreting transmission spectra. JWST solves both of these limitations. For the first time, extremely high precision spectrophotometric light curves spanning the optical to infrared can be obtained within a single transit (e.g. NIRSpec/PRISM 0.5-5.5 $\mu$m), and empirically measured limb darkening for a wide variety of stellar objects can be compared to models. 

In this paper, we use JWST red-optical to near-IR transit data for seven exoplanets with solar-like FGK host stars to empirically compare different limb darkening laws and assess the performance of different atmospheric models. 
In \S \ref{sec:data_reduc} we describe our target selection  data reduction and analysis, in \S \ref{sec:methods} we describe our methods to calculate LD coefficients including a transformation of quadratic coefficient to aid physical interpretation, in \S \ref{sec:results} we describe our results, and in \S \ref{sec:concl} present our conclusions.


\section{Data Reduction and Analysis}\label{sec:data_reduc}
\subsection{Limb-Darkening Optimized Target Selection}

We choose JWST exoplanet targets which have favorable orbits with a low impact parameter, such that the majority of the stellar disc is sampled during transit. To compare targets, we derived the maximum intensity (minimum angle $\theta$) probed for a given exoplanet target as follows.

A point on the stellar disk at projected radial distance $r$ (in units of $R_*$) corresponds to an angle $\theta$ from the sub-observer point, with $r = \sin(\theta)$.
The limb-darkening parameter $\mu$ is related to $r$ by:
\begin{equation}
\mu = \cos(\theta) = \sqrt{1 - \sin^2(\theta)} = \sqrt{1 - r^2}
\end{equation}
At mid-transit, the center of the planet lies at a projected distance $bR_*$ from the center of the star occulting a circular region of the stellar disk. Thus, the point on the planet closest to the stellar center lies at a radial distance $r_{\rm min} = b - p$, where $p = \frac{R_p}{R_*}$.
Substituting $r_{\rm min}$ into the expression for $\mu$ we find,
\begin{equation}
\mu_{\max} = \sqrt{1 - (b - p)^2}.
\label{eq:mumax}
\end{equation}
Thus, for a transiting exoplanet Eq. \ref{eq:mumax} can be used to quickly assess if large areas of the stellar disk are not probed during transit, and therefore not optimal for limb-darkening studies. 
For $\mu_{\max}=0.95$, intensities for about 90$\%$ of the stellar disk are represented during a transit event.

We choose JWST targets with FGK hosts which have been previously observed with either NIRISS/SOSS or NIRSpec/PRISM. The wavelength coverage of NIRISS/SOSS and NIRSpec/PRISM is ideal for limb-darkening studies, as it spans the optical where LD is strong, to the near-IR where LD is weak.  We also choose an initial set of targets to have a $\sim$500 K spacing between stellar $T_{\rm eff}$. In total, we analyzed 7 targets which are listed in Table \ref{tab:fit_comparison}. All of our targets have $\mu_{\max}>0.95$ with three having values $>0.999$. 


\begin{table*}[ht!]
\centering
\caption{JWST transit limb-darkening comparison fits. The minimum BPICS value is bolded, indicating the preferred LD model for each target.}
\begin{adjustbox}{width=8in,right}
\begin{tabular}{lcccc c r c r c r c r}
\toprule
Star, Planet &
$T_{\mathrm{eff}}$ &
$\log(g)$ &
$[{Fe/H}]$ &
$\mu_{\mathrm{max}}$ &
\multicolumn{2}{c}{Linear} &
\multicolumn{2}{c}{Quadratic} &
\multicolumn{2}{c}{3 Parameter} &
\multicolumn{2}{c}{4th Order Nonlinear} \\
\cmidrule(lr){6-7}
\cmidrule(lr){8-9}
\cmidrule(lr){10-11}
\cmidrule(lr){12-13}
&
(K) &
(cgs) &
(dex) &
&
$(R_p/R_*)^2$ (\%) & BPICS &
$(R_p/R_*)^2$ (\%) & BPICS &
$(R_p/R_*)^2$ (\%) & BPICS &
$(R_p/R_*)^2$ (\%) & BPICS \\
\midrule
WASP-121 b & 6770 & 4.25 & 0.17 & 0.9996 &
1.5094 $\pm$ 0.0017 & 1523.52 &
1.4988 $\pm$ 0.0030 & \textbf{1501.95} &
1.4955 $\pm$ 0.0042 & 1503.28 &
1.4905 $\pm$ 0.0061 & 1504.84 \\

WASP-12 A,b & 6380 &4.09 &0.11 & 0.9586 &
1.4423 $\pm$ 0.0011 & 24277.22 &
1.4310 $\pm$ 0.0020 & \textbf{24228.39} &
1.4300 $\pm$ 0.0028 & 24230.48 &
1.4282 $\pm$ 0.0038 & 24232.60 \\


WASP-94 A,b & 6140 &4.16 &0.23 & 0.9703 &
1.1379 $\pm$ 0.0011 & 1563.77 &
1.1267 $\pm$ 0.0016 & 1467.82 &
1.1312 $\pm$ 0.0017 & 1459.59 &
1.1359 $\pm$ 0.0027 & \textbf{1458.67} \\

WASP-6 b & 5380 &4.57 &0.14 & 1.0000 &
2.0834 $\pm$ 0.0020 & 16494.71 &
2.0743 $\pm$ 0.0031 & \textbf{16482.00} &
2.0707 $\pm$ 0.0053 & 16483.99 &
2.0678 $\pm$ 0.0067 & 16485.95 \\

HAT-P-11 b & 4840 &4.58 &0.14 & 0.9573 &
0.3406 $\pm$ 0.0011 & 2206.33 &
0.3374 $\pm$ 0.0014 & \textbf{2196.07} &
0.3365 $\pm$ 0.0014 & 2197.20 &
0.3362 $\pm$ 0.0020 & 2199.73 \\

WASP-107 b & 4425 &4.66 &0.02 & 1.0000 &
2.1126 $\pm$ 0.0026 & 759.09 &
2.0843 $\pm$ 0.0048 & \textbf{699.68} &
2.0915 $\pm$ 0.0065 & 699.87 &
2.0944 $\pm$ 0.0095 & 701.95 \\

WASP-80 b & 4210 &4.65 &-0.25 & 1.0000 &
3.0031 $\pm$ 0.0028 & 1096.02 &
2.9501 $\pm$ 0.0054 & \textbf{936.06} &
2.9542 $\pm$ 0.0086 & 938.18 &
2.9525 $\pm$ 0.0104 & 940.31 \\
\midrule
\multicolumn{5}{l}{\textit{Aggregate (all 7 targets), $\Delta$BPICS:}} &
\multicolumn{2}{c}{+408.69} &
\multicolumn{2}{c}{\textbf{0.00}} &
\multicolumn{2}{c}{+0.62} &
\multicolumn{2}{c}{+12.08} \\
\multicolumn{5}{l}{\textit{Aggregate (excl. WASP-94A), $\Delta$BPICS:}} &
\multicolumn{2}{c}{+312.74} &
\multicolumn{2}{c}{\textbf{0.00}} &
\multicolumn{2}{c}{+8.85} &
\multicolumn{2}{c}{+21.23} \\
\bottomrule
\end{tabular}
\end{adjustbox}
\label{tab:fit_comparison}
\end{table*}


\begin{table}[ht!]
\caption{Observations. The JWST ID numbers associated with the datasets used are listed, along with stellar activity indices and orbital parameter references for the target sample.}
\label{tab:stellar_activity}
\begin{tabular}{lclc}
\toprule
Name     & ID \# & $\log R'_{\rm HK}$ & $P$, $e$, $\omega$ Ref. \\
\midrule
WASP-121 & 1201 & $-4.87$\textsuperscript{1}  & 4 \\
WASP-12A & 5924 & $-5.500$\textsuperscript{2} & 5 \\
WASP-94A & 5924 & $-5.18$\textsuperscript{3}  & 6 \\
WASP-6   & 5924 & $-4.741$\textsuperscript{2} & 7 \\
HAT-P-11 & 5924 & --       & 8 \\
WASP-107 & 1201 & --       & 9 \\
WASP-80  & 5924 & --       & 6 \\
\bottomrule
\end{tabular}
\tablecomments{References: 
(1)~\cite{Borsa2021}; 
(2)~\cite{sing2016_nature}; 
(3)~\cite{2022MNRAS.510.4857A}; 
(4)~\cite{2024AJ....168..231S}; 
(5)~\cite{2025ApJ...993...78S}; 
(6)~\cite{2023ApJS..265....4K}; 
(7)~\cite{2022ApJS..259...62I}; 
(8)~\cite{Basilicata2024}; 
(9)~\cite{2024NatAs...8.1562M}.}
\label{tab:obs}
\end{table}

\subsection{\texttt{FIREFLy} Data Reduction and Light-curve fitting}
The targets for our study are listed in table \ref{tab:obs}. We used NIRISS/SOSS transit data of
WASP-121~b \citep{2025NatCo..1610822A,2026A&A...706A...2P} from program GO-1201 (PI. Lafreniere, D.) as well as NIRISS/SOSS data of
WASP-107~b \citep{2026NatAs..10..258K}
from GO-1201 (P.I. Lafreniere. D.). 
For the rest of our targets, we used data from the JWST Grand Tour program (GO-5924, PI Sing) including:
WASP-94~Ab \citep{2025arXiv250510910M},
WASP-6~b (Fu et al. 2026, in prep.),
HAT-P-11~b (Bennett, et al. 2026, in prep.), 
WASP-12~b (Schmidt et al. 2026 in prep.),
and WASP-80~b (Gascon et al. 2026, in prep.). 

All of the data were handled and reduced similarly with the same software to produce a homogenous dataset. We used the Fast InfraRed Exoplanet Fitting Lyghtcurve (\texttt{FIREFLy}) suite to reduce the JWST datasets  \citep{Rustamkulov2022ApJ...928L...7R, Rustamkulov2023Natur.614..659R, 2024Natur.630..831S} with additional details for NIRISS/SOSS found in \cite{2025arXiv251116771W}. The reductions start from the uncalibrated \texttt{uncal.fits} files and use the JWST Calibration pipeline with customizations to optimize the time series observation data reduction. These customized routines include removal of 1/$f$ noise both at the group and integration level, bad pixel flagging and removal, cosmic ray and snowball cleaning both spatially and from the time series, measuring spectral trace drifts with cross correlation, a customized background subtraction for SOSS, and an optimized extraction of the stellar spectrum. 

For all targets and data, we used a wavelength range of 0.85 to 2.75 $\mu$m (which covers SOSS order 1) to fit white light curves. Several SOSS targets had significant contamination in order 2, so those wavelengths were excluded in this study. We used the transit model of \cite{2015PASP..127.1161K}, with free parameters of the mid-transit time ($T_0$), the planet-to-star radius ratio ($R_p/R_*$), the semi-major axis in units of stellar radii ($a/R_*$), impact parameter ($b$), and baseline flux which varied with a visit-long slope. The periods and eccentricity values were set fixed to literature values see Table \ref{tab:obs}). The white light curves were freely fit with four different limb darkening parameterizations: a linear, a quadratic, three parameter, and 4-parameter non-linear limb darkening law. For each LD parameterization, the coefficients were freely fit. Unlike the other laws, for the 4-parameter non-linear law, we additionally set uniform priors on the coefficients restricting their range to between -2 and 2. This prior keeps the solutions within the general range of the model predictions helping to avoid overly unphysical LD profiles. We checked the prior did not significantly effect the results for the non-linear law. For both WASP-6 and WASP-12A, we found the prior had the effect of cutting off the tails of the posterior distribution and help prevent alternate non-physical solutions. While beyond the scope of this work, we encourage future studies to adopt more physically motivated priors when fitting the 4-parameter non-linear law. The results can be found in Table \ref{tab:fit_comparison}. For  model selection, we also measured the Bayesian Predictive Information Criterion Simplified (BPICS) statistic for each fit following the recommendation of \cite{2026ApJS..283...10T}. The BPICS is a model selection criterion that weighs the goodness-of-fit with the number of free parameters and is designed to select the model with the best predictive power. Unlike other information criteria such as the BIC or AIC, it does not rely on point estimates using the whole posterior instead, and is insensitive to priors and binning choices. Its penalty term matches the AIC, though higher-dimensional fits will also have more posterior mass at lower probabilities. 
For each target, we also fit for the quadratic limb darkening coefficients in the spectroscopic channels, fixing the system parameters to the white-light curve best-fit values. 

\subsection{Limb-darkening Law Selection and Transit-depth Biases}

When comparing the values in Table \ref{tab:fit_comparison}, for each individual target the statistics between the higher-order limb darkening laws are not often definitive, with only the linear law clearly disfavored in all cases. In most cases, we find the quadratic law usually has a better fit and marginally lower BPICS value with a median $\Delta$BPICS of 1.3 and 3.7 when compared to the 3-parameter and 4-parameter respectively. When aggregating the statistics, clearer trends emerge. 
We also compute an aggregate comparison by summing the BPICS values across targets and comparing the resulting $\Delta$BPICS values.
Summing the BPICS values for all seven stars, the 3-parameter law provides equally good fits as the quadratic which has a slightly lower BPICS value by $\Delta \mathrm{BPICS}=0.62$, corresponding to posterior odds of only 1.4:1 (57.6\% probability) in favor of the quadratic law assuming equal prior probabilities. In contrast, the 4-parameter law is confidently disfavored, with $\Delta \mathrm{BPICS}=12.1$, corresponding to odds of 420:1 (99.76\% probability) against the 4-parameter model. 
Following the recommendations of  \cite{2026ApJS..283...10T}, we report these results as odds ratios and posterior model probabilities rather than converting them to Gaussian-equivalent detection significances. 
Six of the seven stars favor a quadratic LD law. Excluding WASP-94A, the BPICS statistic clearly indicates a quadratic law is preferred, with the 3-parameter law having a $\Delta \mathrm{BPICS}=8.85$ (odds ratio 83:1, 98.8\%) higher and 4-parameter 21.23 higher (odds ratio 4$\times10^4$:1).  For the late F-dwarf WASP-94A ($T_{eff}$=6140 K), higher orders are preferred with the 3-parameter and 4-parameter having nearly equal fits $\Delta \mathrm{BPICS}<1$ while the quadratic is disfavored $\Delta \mathrm{BPICS}=9.1$. Thus, taken in aggregate, the quadratic and 3-parameter LD laws offer good fits to the JWST transit light curves while the 4-parameter law is disfavored. For most stellar types, the quadratic LD law is preferred.

\begin{figure}[ht!]
\centering
\includegraphics[width=\linewidth]{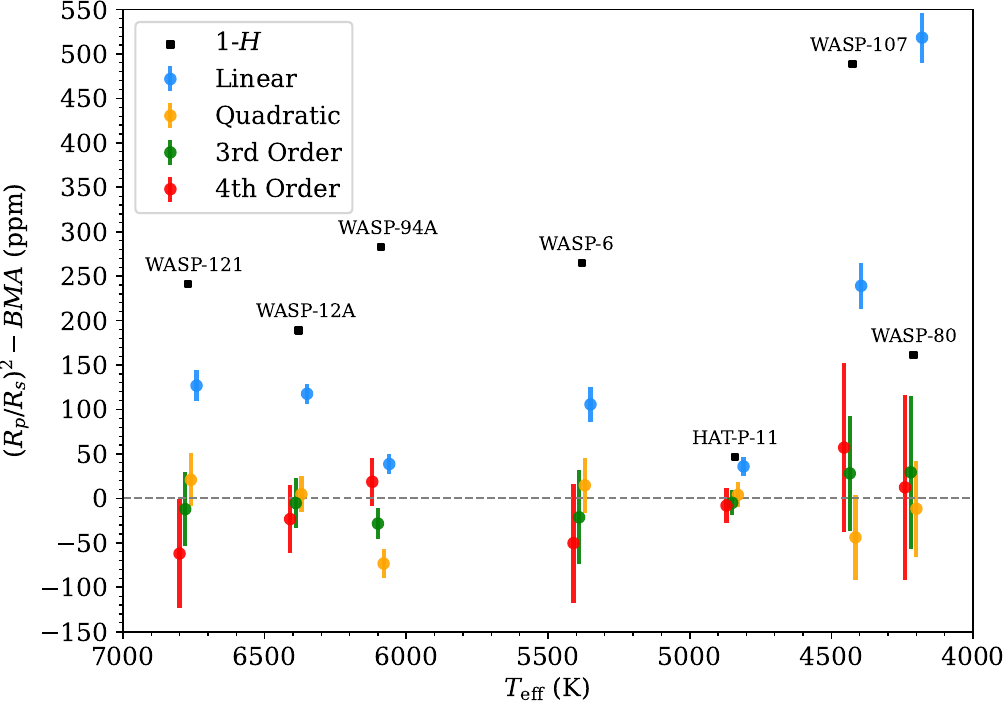}
\vspace{-0.5cm}
\caption{
Fit planetary transit depths from JWST white light curves using different limb-darkening assumptions as a function of stellar effective temperature. The measurements have been slightly offset in temperature for clarity. The depths are relative to the Bayesian Model Average (BMA) transit depth, using the BPICS to measure the model probabilities. 
The 1$\sigma$ uncertainties are shown as well as the transmission spectral amplitude of one pressure scale height, $H$. 
}
\label{fig:modelavg_radii}
\end{figure}

Given these different limb darkening laws, a central question is how does each LD law bias the derived $R_p/R_*$ value? To compare the fit values while accounting for the uncertainty in model selection, we calculated the Bayesian model average (BMA) $R_p/R_*$ using all four LD models. Each LD-model fit $R_p/R_*$ value was then compared to the BMA (see Figure \ref{fig:modelavg_radii}). For scale, we also compared these values with the transmission spectral amplitudes, estimated assuming 1 pressure scale height ($H$) features. As seen in Fig. \ref{fig:modelavg_radii}, the linear LD model is associated with the largest bias in transit depth, with biases of over $>100$ ppm seen. However, the linear model is confidently disfavored over higher-order LD laws for all targets and is generally not used in the literature to fit transit light-curves. Comparing the higher order LD laws, in six of our seven targets the fit radii are all in excellent agreement ($<1-\sigma$ difference) both between themselves and the BMA. The quadratic law for 6/7 targets gives radii with a standard deviation of 14 ppm relative to the BMA value, indicating a small potential bias which is well below the amplitude of the $\sim$200 ppm typical 1-$H$ transmission spectral features. For WASP-94A, the quadratic is 73 ppm deviant from the BMA, though this bias is still small at 1/4-$H$. 

\begin{figure}[t!]
\centering
\includegraphics[width=\linewidth]{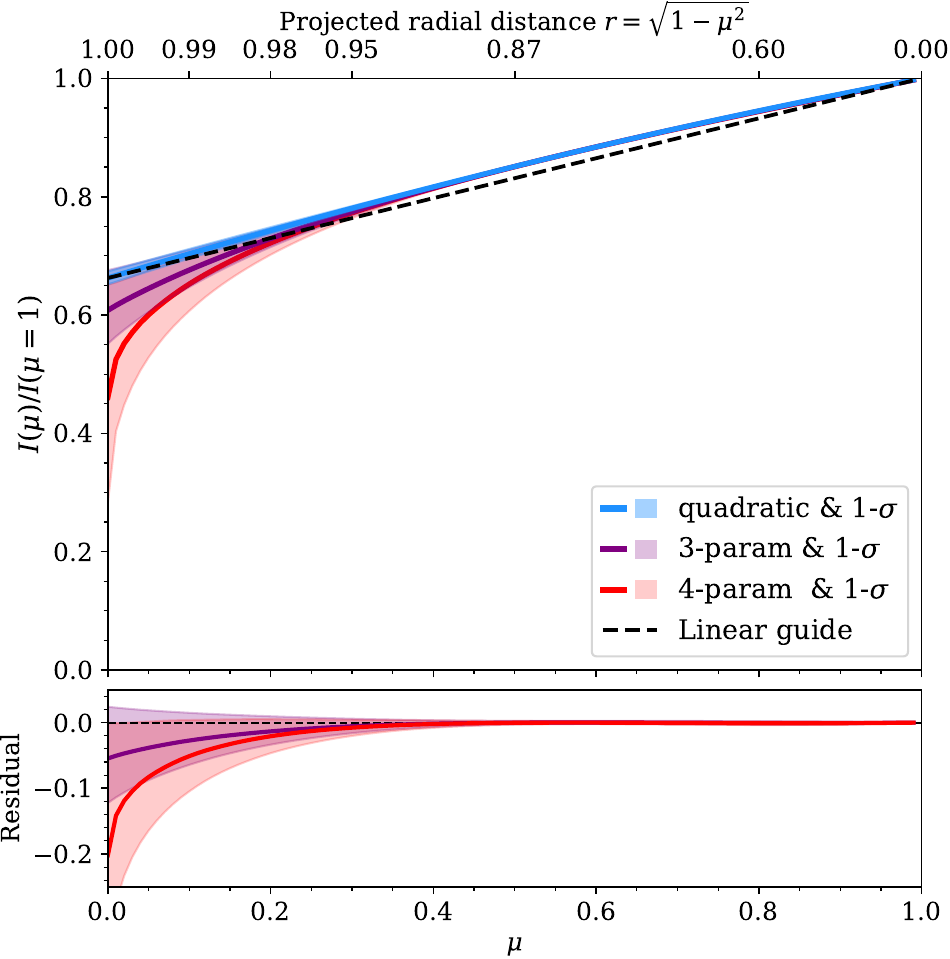}
\vspace{-0.5cm}
\caption{Posterior intensity profiles $I(\mu)/I(\mu=1)$ from the JWST NIRSpec/Prism white-light-curve fit of WASP-6. The quadratic (blue), 3-parameter (purple), and 4-parameter (red) 
limb-darkening laws are shown with their $1-\sigma$ posterior envelopes. A linear guide (dashed black) connects the disk center to the best-fit limb intensity. \textit{Bottom:} Residuals of the 3-parameter and 4-parameter intensity compared to the best-fit quadratic law. The parameterizations only differ substantially below $\mu\lesssim0.2$ which represents only about 4\% of the stellar disk area.}
\label{fig:w6_i}
\end{figure}

Inspection of the retrieved intensity profiles for the different LD parameterizations help illustrate why the different laws give largely consistent transit depths. 
As seen in Fig.~\ref{fig:w6_i}, a quadratic LD law gives nearly the same retrieved intensity profile as higher-order laws for the vast majority of the stellar disk, with deviations of $<1\%$ in intensity for $\mu>0.3$. While deviations larger than $1\%$ occur for $\mu<0.3$, this region only covers the outer 5\% of the stellar-projected radial distance. Thus the main difference is only at the very outer limb, which contributes minimally to the overall flux, and even high-SNR JWST data struggle to detect any high-order intensity drop at the limb. Limb-darkening laws designed to reproduce model intensities at the very limb (e.g.\ the 4-parameter non-linear law, \citealt{Claret2000}) can overfit the outer $\sim$1\% of radial distance, given lower-order functions are statistically preferred, while the bulk of the intensity profile that determines the stellar $I$--$\mu$--$r$ relation is well-captured by simpler laws, with $\mu<0.2$ only representing 4\% of the stellar disk area. The key result is not that higher-order laws are incorrect, but that they are unnecessary. The additional parameters are poorly constrained by the data and do not improve the fit, indicating that any intensity structure at the extreme limb does not measurably affect transit depths at JWST precision.

From this analysis, we generally recommend fitting the LD with a quadratic law, statistically checking the quality of fit and transit depth vs. a 3-parameter law. The quadratic law is found to be statistically preferred over higher order LD laws, and offers a minimal transit depth bias. The computational speed of the quadratic vs higher order laws also makes it a very pragmatic choice.


\section{Methods} \label{sec:methods}
To compare our empirically derived LD fits to models, we calculated limb darkening model coefficients for each of our target stars, using \texttt{Exotic-LD} \citep{2024JOSS....9.6816G}. We calculated LD coefficients for the PHOENIX, Stagger, and MPS-ATLAS grids for both sets 1 and 2 \citep{husser2013new,magic2015stagger,kostogryz2022stellar,kostogryz2023mps}. These models have different underlying geometries, as the MPS-ATLAS grids are 1D plane-parallel models, while the PHOENIX model is a 1D model with spherical geometry, while the Stagger models are 3D. The minimum value of $\mu$ used in the intensity calculation ($\mu_{\rm min}$) was varied, with coefficients derived for both the wavelength ranges of the JWST NIRISS/SOSS and NIRSpec/PRISM
white light curves and spectroscopic bins. As detailed in \S \ref{sec:phoex}, the spherical PHOENIX models required unique handling, including $\mu$ angles being rescaled using a critical angle ($\mu_{\rm cri}$) which defines the location of the stellar limb.

The stellar parameters for the effective temperature $T_{\mathrm{eff}}$, surface gravity $\log(g)$, and metallicity $[{Fe/H}]$ used in this study are given in Table \ref{tab:fit_comparison}. Our study focused on stars spanning a range $T_{\mathrm{eff}}$, as this is the first major stellar parameter where JWST transit data are readily available across a range of values. 
These parameters have been derived using the \texttt{isochrones} \citep{mor15} package with a simultaneous Bayesian
fit of the MESA Isochrones \& Stellar Tracks (MIST) isochrone grid
\citep{pax11,pax13,pax18,pax19,jer23,dot16,cho16} to a curated collection
of photometric, spectroscopic and GAIA \citep{gai16,gai18} parallax data for each star (\citealt{2024AJ....168..231S}, McCreery et al. in prep.).

\subsection{Spherical PHOENIX model calculations}\label{sec:phoex}

Stellar atmosphere models that calculate radiative transfer in spherical geometry are known to exhibit angular intensity profiles with very low intensity at small $\mu$. This effect arises because rays passing through the upper atmospheric layers must be included in order to compute the flux at each depth and converge to radiative equilibrium. 

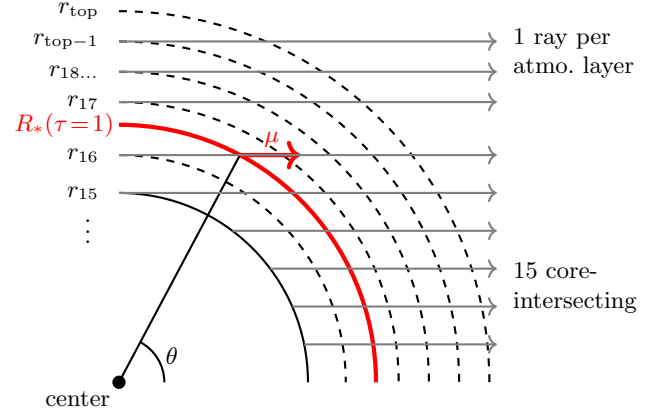
\begin{figure}[t]
\centering
\begin{tikzpicture}[scale=1.0]
%
%
\draw[thick] (0,0) ++(90:2.5) arc (90:0:2.5);
\draw[thick, dashed] (0,0) ++(90:3.0) arc (90:0:3.0);
%
\draw[ultra thick, red] (0,0) ++(90:3.4) arc (90:0:3.4);
%
\draw[thick, dashed] (0,0) ++(90:3.7) arc (90:0:3.7);
\draw[thick, dashed] (0,0) ++(90:4.1) arc (90:0:4.1);
\draw[thick, dashed] (0,0) ++(90:4.5) arc (90:0:4.5);
\draw[thick, dashed] (0,0) ++(90:4.9) arc (90:0:4.9);
%
%
\draw[thick] (0,0) -- (1.6, 3.0);
%
\draw[->, ultra thick, red] (1.6, 3.0) -- (2.4, 3.0);
\node[red, above] at (2.025, 3.0) {$\mu$};

\draw[thick] (0.6, 0) arc (0:61.93:0.6);
%
\node at (0.7, 0.35) {$\theta$};

\node[left] at (-0.15, 2.5) {$r_{15}$};
\node[left] at (-0.15, 3.0) {$r_{16}$};
\node[left] at (-0.15, 3.7) {$r_{17}$};
\node[left] at (-0.15, 4.1) {$r_{18...}$};
\node[left] at (-0.15, 4.5) {$r_{\rm top-1}$};
\node[left] at (-0.15, 4.9) {$r_{\rm top}$};
%
\node[red, left] at (0.05, 3.4) {$R_*(\tau\!=\!1)$};
%
\node[left] at (-0.25, 2.1) {$\vdots$};
%
\foreach \r in {2.5, 3.0, 3.7, 4.1, 4.5} {%
    \draw[->, thick, gray] (0, \r) -- (5.0, \r);%
}
%
\node[right, align=left, font=\small] at (5.1, 4.35) {1 ray per\\atmo.\ layer};
%
\foreach \y in {0.5, 1.0, 1.5, 2.0} {%
    \pgfmathsetmacro{\xstart}{sqrt(2.5*2.5 - \y*\y)}%
    \draw[->, thick, gray] (\xstart, \y) -- (5.0, \y);%
}
%
\node[right, align=left, font=\small] at (5.1, 1.25) {15 core-\\intersecting};
%
\fill (0,0) circle (2.5pt);
\node[below left] at (0,0) {center};
\end{tikzpicture}
\vspace{-0.5cm}
\caption{Geometry of ray paths in spherical PHOENIX atmosphere models (not to scale). 
Dashed arcs above $R_*$ represent the modeled extended atmosphere grid. 
The PHOENIX spherical model grid runs from the center of the star with 15 rays intersecting 
an optically thick core and includes a modeled atmosphere (64 layers) out to low $\sim$nbar pressures 
at the top, $r_{\rm top}$. The thick red arc illustrates the effective stellar radius at 
the photosphere, $R_*(\tau=1)$. For limb darkening, $\mu=\cos(\theta)$ needs to be re-normalized so 
that $\mu=0$ corresponds to $R_*(\tau=1)$.}
\label{fig:phoenix_rays}
\end{figure}

In PHOENIX, spherical models are commonly computed on a 64-layer atmosphere grid \citep[e.g.][]{husser2013new}, see Figure \ref{fig:phoenix_rays}, with one tangential ray assigned to each layer except the outermost, giving 63 tangential rays, plus 15 core-intersecting rays to sample the flux emerging from below (see Fig. \ref{fig:phoenix_rays}). These rays sample the low- and high-$\mu$ regimes, respectively, while the atmosphere itself extends to an optically thin top-of-the-atmosphere well above the star’s optical depth $\tau\approx1$ photosphere \citep[e.g.][]{2008A&A...491..633L,2018A&A...618A..20C}. In the \cite{2013A&A...553A...6H} grid, for example, the model extends to a top radius $r_{\rm top}$ near nanobar pressures. As a result, the tangential rays begin probing the optically thin upper atmosphere at small $\mu$ ($\mu \approx 0.05$), where the intensities fall by orders of magnitude but do not reach zero. For transit light curves, however, the stellar limb is conventionally defined at the edge of the optically thick photosphere, since ingress begins only once the planet occults those rays. The raw $\mu$ values in spherical atmosphere models must therefore be re-scaled so that $\mu=0$ marks the edge of the optically thick emitting disk rather than the top of the model atmosphere. We therefore explore two rescaling prescriptions for spherical PHOENIX intensities: a numerical max-derivative method and a physically motivated method that defines the effective limb at the $\tau_{\rm slant}=1$ surface. This correction is important because \texttt{Exotic-LD} \citep{2024JOSS....9.6816G} does not apply it natively and will otherwise over-predict limb darkening, even when fitting only intensities above a cutoff in $\mu$.


\subsubsection{Numerical $\mu$-rescaling method}
\begin{figure}[t!]
\centering
\includegraphics[width=0.95\linewidth]{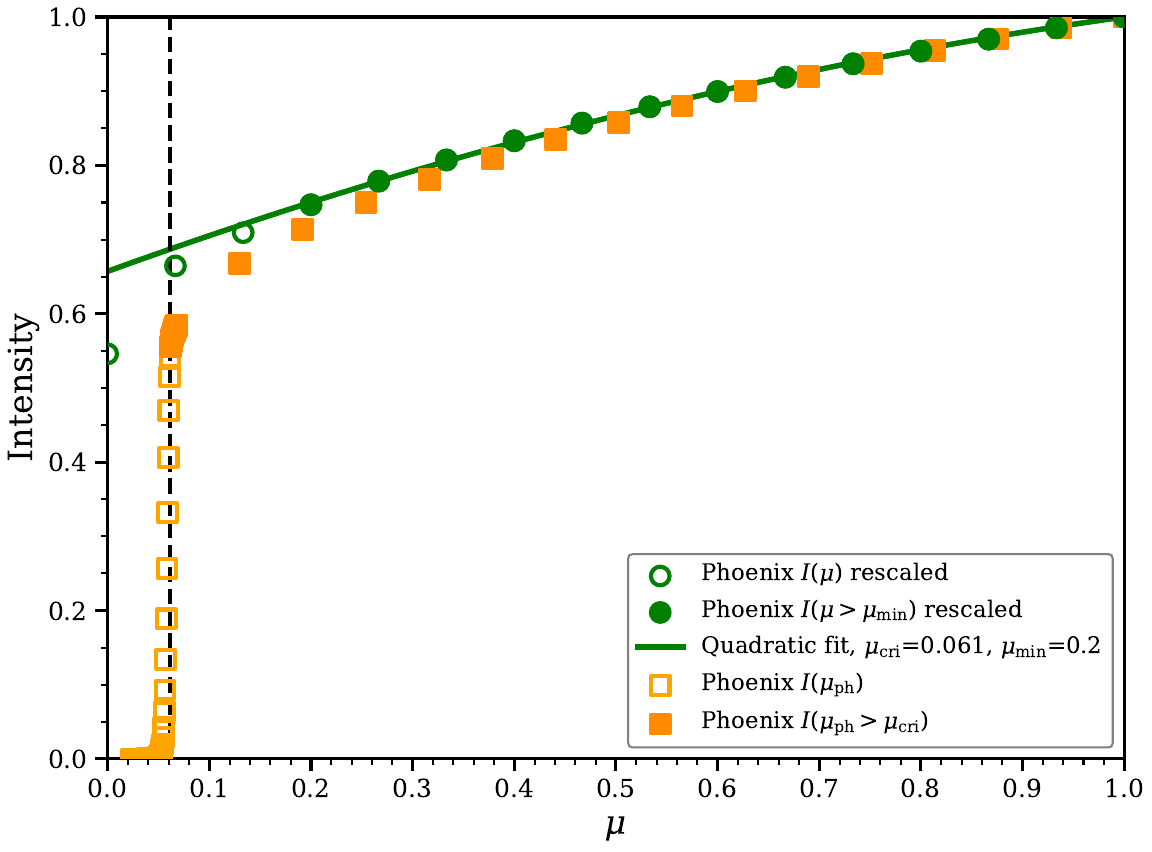}
\includegraphics[width=1\linewidth]{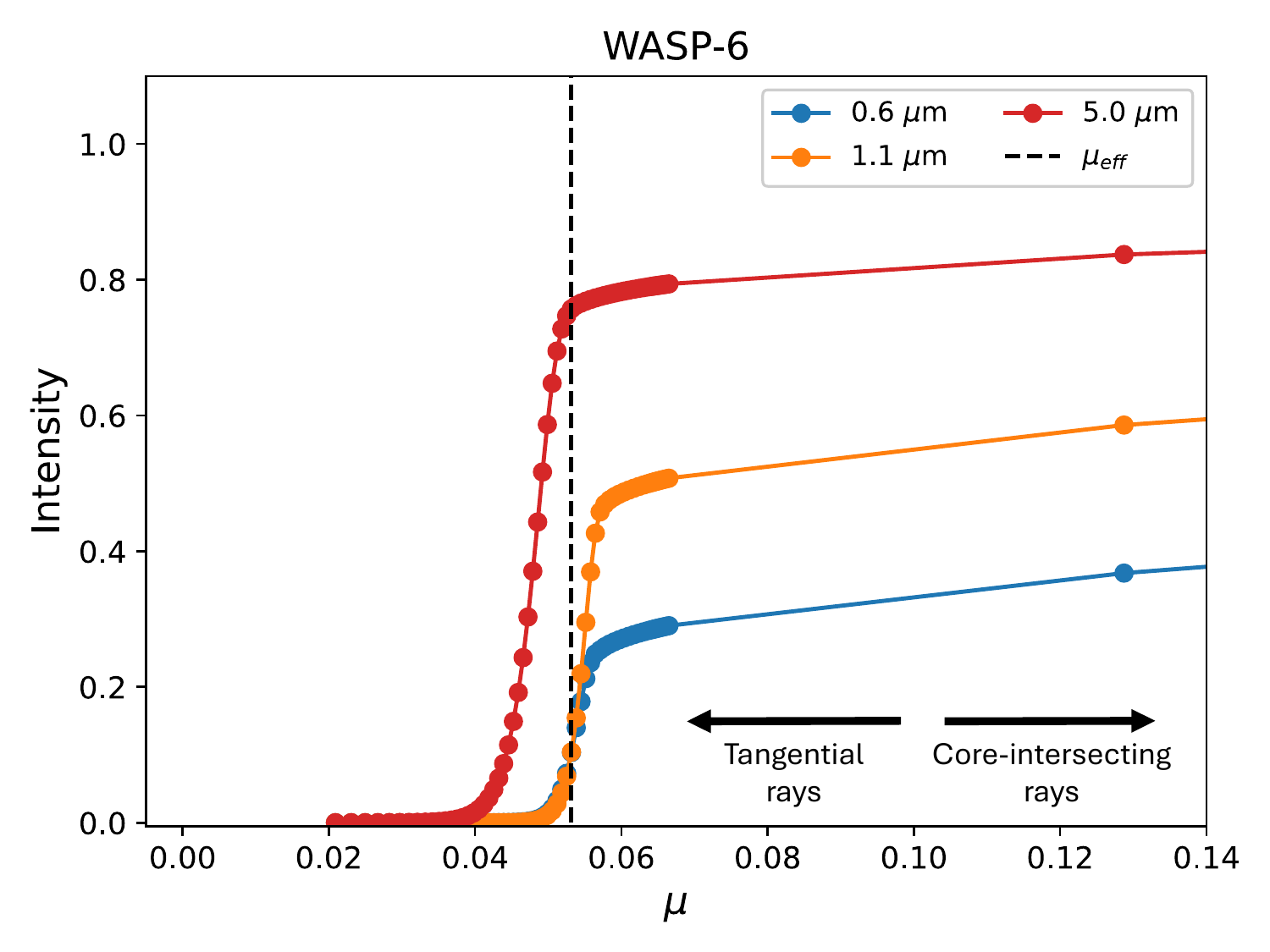}
\vspace{-0.75cm}
\caption{Spherical PHOENIX intensity profiles for WASP-6 
($T_{\rm eff} = 5400$~K, $\log g = 4.5$). 
\textit{Top:} Illustration of the maximum-derivative $\mu$-rescaling procedure.  
The stellar limb is redefined at $\mu_{\rm cri}$ (determined from the maximum 
$dI/d\mu$ derivative), and the remaining intensities (solid) are rescaled and interpolated 
onto a uniform $\mu$ grid. An optional $\mu_{\rm min}$ cutoff can be applied when 
fitting limb-darkening laws to avoid residual edge effects.
\textit{Bottom:} Illustration of the $\tau=1$- $\mu$-rescaling method. Intensity profiles at optical, near-IR, and mid-IR wavelengths 
showing the transition between the densely sampled tangential rays (low $\mu$) and 
core-intersecting rays (high $\mu$). The vertical dotted lines indicate $\mu_{\rm eff}$, 
calculated from the $\tau_{\rm slant}=1$ surface as described in Section~\ref{sec:phoex}.
}
\label{fig:W6_rescale}
\end{figure}

We first used a variant of the max-derivative method to rescale the original Phoenix $\mu$ values, $\mu_{\rm ph}$. The intensities $I(\mu_{\rm ph})$ were calculated
for each wavelength bin of interest using \texttt{Exotic-LD} \citep{2024JOSS....9.6816G}. The angle corresponding to the maximum $dI/d\mu$ derivative for that wavelength bin was then determined numerically. While other studies have re-scaled using this angle as the $\mu_{\rm cri}$ limb \citep{2015MNRAS.450.1879E,2015MNRAS.453.3821P,2018A&A...618A..20C,2023MNRAS.519.3723M}, we also elected to remove the next four larger adjacent $\mu_{\rm ph}$ grid points, such that the majority of the high-derivative $dI$/$d\mu$ region was removed.  We then re-scaled the remaining $\mu_{\rm ph}$ values as $\mu=(\mu_{\rm ph}-\mu_{\rm cri})/(1-\mu_{\rm cri})$
before interpolating $I$ on a constant$-\mu$ grid. The limb-darkening calculations then used these rescaled-interpolated set of $I(\mu)$ to derive coefficients, with an optional $\mu_{\rm min}$ also included (see Fig. \ref{fig:W6_rescale}).  
For WASP-6, we found a value of $\mu_{\rm cri}$=0.061 (see. Fig. \ref{fig:W6_rescale}), which agrees with the value of $\mu_{\rm cri}$=0.0548$\pm$0.0055 from \cite{2023MNRAS.519.3723M}  which was derived for a $T_{\rm eff}$ 6000 K star.

\subsubsection{Physically motivated $\tau=1$ $\mu$-rescaling method}\label{sec:mueff}
We also explored a physically motivated method to rescale $\mu_{\rm ph}$ to account for the difference between $\mu = 0$ in the model versus $\mu$ at the edge of the optically thick disk. We can compute the $\mu$ value of the first optically-thick tangential ray, $\mu_{\rm eff}$, and rescale the intensity profile to that point. This will correspond to the ray where the slant optical depth, $\tau_{\rm slant}$ is unity, given by 

\begin{equation}
    \tau_{\rm{slant},}{_i} = \tau_{\rm vertical,}{_i}/\sqrt{2\pi R_i/H_i}
\end{equation}

\noindent where $\tau_{\rm vertical}$ is the vertical optical depth of layer $i$, $R_i$ is the radius of the star, and $H$ is the stellar atmospheric scale height. Here, $\tau_{\rm vertical}$ is PHOENIX's standard optical depth grid defined at 0.5\,$\mu$m if $T_{\rm eff}\geq5000$ K and 1.2\,$\mu$m if $T_{\rm eff}<5000$ K \citep{husser2013new}.

$\mu_{\rm eff}$ will be different for each model, depending on the atmospheric structure. To post-process the \cite{husser2013new} models to estimate $\mu_{\rm eff}$, we must calculate the radius grid of the model which can be done by reconstructing the solution to hydrostatic equilibrium solution using the vertical optical depth, temperature, pressure, and mass density tabulated in the ``ATMOS.fits" outputs of the models.

\begin{figure}
    \centering
    \includegraphics[width=1\linewidth]{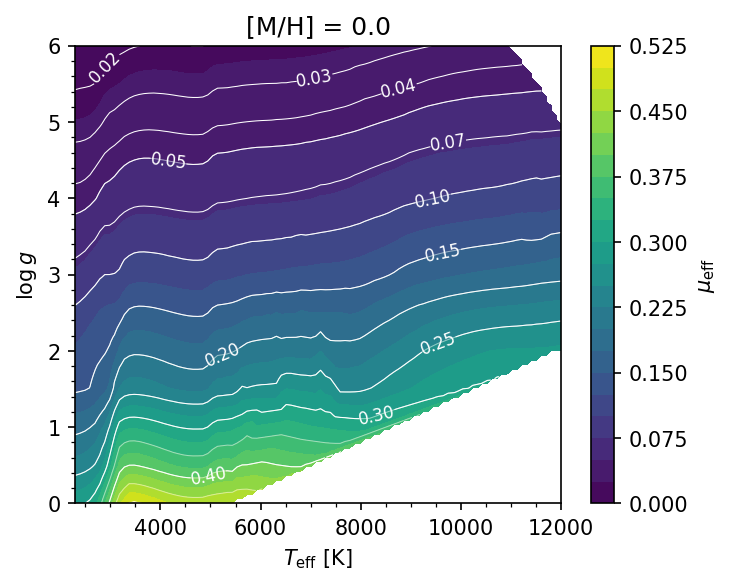}
    \vspace{-0.75cm}
    \caption{Effective limb angle $\mu_{\rm eff}$ as a function of stellar parameters 
for the \cite{husser2013new} PHOENIX spherical model grid at solar metallicity ([M/H] = 0.0). 
$\mu_{\rm eff}$ represents the $\mu$ value at which the slant optical depth reaches unity. Cooler stars and lower 
surface gravities produce smaller $\mu_{\rm eff}$ values, corresponding to more extended 
atmospheres where the $\tau_{\rm slant}=1$ surface lies closer to the model's outer boundary. 
Contour labels indicate $\mu_{\rm eff}$ values.}
    
    \label{fig:phx_mueff} 
\end{figure}


We numerically reconstruct the radius at each level using hydrostatic equilibrium, building outward from the $\tau = 1$ level in the model, where the model-output effective radius, $r_{\rm eff}$, is defined:

\begin{equation}
    r_{i} = r_{i-1}+dr_i
\end{equation}
\begin{equation}
    dr_i = -(P_i - P_{i-1})/ (\rho(r_i) g(r_i))
\end{equation}

\noindent With a reconstructed radius grid, we can then calculate the $\mu$ at the $\tau_{\rm slant}=1$ level to be

\begin{equation}
    \mu_{\rm eff} = \sqrt{1 - (r_{\rm thick}/r_{\rm top})^2},
\end{equation}

\noindent where $r_{\rm thick}$ is the radius of the layer where $\tau_{\rm slant} = 1$ and $r_{\rm top}$ is the radius at the top of the model.

We calculate the $\mu_{\rm eff}$ value for all models in the \cite{husser2013new} grid (see Fig. \ref{fig:phx_mueff}). 
The complete list of computed $\mu_{\rm eff}$ values for the PHOENIX spherical atmosphere grid is provided as a machine-readable table, with an excerpt shown in Table~\ref{tab:mueff}.
Some low-gravity giant star models did not have well-defined density at low pressures so these were skipped. 

Figure~\ref{fig:W6_rescale} shows an example of the $\mu$ values for WASP-6. The $\mu_{\rm eff}$ value lies near the beginning of the decrease in intensity, providing an ideal place to rescale for limb-darkening coefficient fitting using common laws. For WASP-6, $\mu_{\rm eff}$=0.0531 which is close to the value of $\mu_{\rm cri}$=0.061 when using the numerical method. One drawback of our method here is that it the standard PHOENIX optical depth grid, which is calculated only at one wavelength (either 0.5 or 1.2 $\mu$m) to set the reference vertical structure. At longer wavelengths, where limb darkening is not as strong, the cutoff may be at somewhat larger $\mu$ than where the atmosphere becomes optically thick at IR wavelengths. Future model grids tailored to exoplanet limb-darkening calculations may wish to use the method above to define $\mu_{\rm eff}$, but using the specific optical depth for each wavelength.

\subsection{The Plane-Parallel Approximation, Spherical Corrections and Limb Angle Cutoffs}
Previous studies have demonstrated that plane-parallel (e.g. MPS-ATLAS) and spherical atmospheric models (e.g. PHOENIX) predict significantly different center-to-limb intensities near the stellar edge \citep{2003A&A...412..241C,2013A&A...556A..86N,2013A&A...554A..98N}.
Here we provide a simple analytic estimate showing that this discrepancy naturally arises from the divergent optical path length in the plane-parallel approximation as $\mu\rightarrow0$.
When calculating limb-darkening coefficients using 1D plane-parallel atmospheric models, a common practice is to exclude the intensities closest to the limb, which often yields better agreement with observational data (e.g. \citealt{Sing2009,2010A&A...510A..21S,2015MNRAS.450.1879E,2024JOSS....9.6816G}).
We motivate this practice geometrically.

At angles very close to the limb, the optical depth in a plane-parallel (PP) atmosphere becomes systematically larger than in a more realistic spherical geometry, causing the emergent intensities to be underestimated.
We estimate the angle at which the plane-parallel approximation begins to break down as follows.
We assume a hydrostatic atmosphere with pressure scale height $H$, constant opacity $\kappa$, and density profile $\rho(z)=\rho_0 e^{-z/H}$, where $\rho_0$ is the density at the reference radius $R_{\star}$.
In PP geometry, the optical depth for a ray traveling at 
angle $\mu=\cos(\theta)$ from the normal is:
\begin{align}
\tau_{PP} 
    = \frac{\kappa \rho_0}{\mu} \int_0^\infty e^{-z/H} \, dz 
    = \frac{\kappa \rho_0 H}{\mu}
    = \frac{\tau_0}{\mu}
\end{align}
where $\tau_0 \equiv \kappa \rho_0 H$ is the vertical optical depth scale. Since $\mu$ is the direction of the ray, it is constant along the path in PP geometry and factors out of the integral. In spherical geometry, however, the angle between the ray and the local radial direction changes along the path, giving instead:
\begin{align}
\tau_{sph} = \kappa \rho_0 \int_0^\infty e^{-z/H} \frac{(R_* + z) \, dz}{\sqrt{(R_* + z)^2 - p^2}}
\end{align}
with $p=R_*\sqrt{1-\mu^2}$. After substituting $x=z/H$,  simplifying, and assuming $H \ll R_*$, we find
\begin{align}
\tau_{sph} \approx \frac{\tau_0}{\mu} \int_0^\infty \frac{e^{-x}}{\sqrt{1 + \alpha x}} \, dx
\end{align}
where $\alpha=\frac{2H}{R_* \mu^2}$.
The ratio of optical depths is then,
\begin{align}
\tau_{sph}/\tau_{PP} &= \sqrt{\frac{\pi}{\alpha}} \cdot e^{1/\alpha} \cdot \mathrm{erfc}\left(\frac{1}{\sqrt{\alpha}}\right)
\label{eq:geocor}
\end{align}
which uses the upper incomplete gamma function 
$\Gamma(1/2, a) = \sqrt{\pi}\,\mathrm{erfc}(\sqrt{a})$.
Assuming $H/R_{\star}\ll1$, we define a geometric correction factor $f_\mu\equiv\frac{\tau_{\rm sph}} 
{\tau_{\rm PP}}$, 
with a second-order approximation for EQ. \ref{eq:geocor} good for $\mu>0.02$,
\begin{align}
f_\mu \approx 1 - \frac{H}{R_* \, \mu^2} + \frac{3 H^2}{R_*^2 \, \mu^4}.
\label{Eq_geo}
\end{align}

From Eq. \ref{Eq_geo}, for the Sun we find the optical depth for plane-parallel models is 8.2\% 
too large at $\mu=0.05$ and still 2.1\% too large at $\mu=0.1$. 
A more realistic 
atmosphere with decreasing temperature, changing opacity with altitude, and magnetic fields may further increase 
this correction. 
The larger PP optical depth causes rays at small $\mu$ to sample higher, cooler atmospheric layers than would occur in a spherical atmosphere, leading to artificially low emergent limb intensities.
We note that inspection of the 1D PP atmospheric models typically do indeed show the intensity values falling rapidly at the very limb, consistent with this geometric explanation.

This overestimation supports the removal of the intensities closest to the limb when calculating limb-darkening coefficients in 1D PP models. 3D hydrodynamic models in a cartesian box (e.g. \citealt{magic2015stagger}) may also be affected. In this work, we adopt a conservative cutoff of $\mu_{\rm min}\ge0.1$ for sun-like stars, motivated in part by these geometric considerations and by previous empirical studies showing improved agreement with observations when the smallest-$\mu$ intensities are excluded.



\subsection{Limb Darkening Transformations}

To facilitate a better physical understanding when analysing the  limb darkening coefficients, we re-parameterized the quadratic limb darkening law,
\begin{align}
I(\mu) &= 1 - c_1(1 - \mu) - c_2(1 - \mu)^2, 
\end{align}
in terms of two parameters, $\ell$ and $\delta$. We define $\ell$ as the intensity at the limb ($\mu = 0$):
\begin{align}
\ell &=1 - ( c_1 + c_2).
\end{align}
$\ell$ sets the overall slope between the limb and disk center. The linear law with $\ell$ is then,
\begin{align}
I_{\rm lin}=\ell+(1-\ell)\mu.
\end{align}
We also define $\delta$ as the nonlinearity at $\mu$=1/2 relative to the linear law, which measures the drop/rise of intensity compared the linear projection to the limb. Thus, $\delta$=0 gives a linear law while non-zero values introduce curvature, with larger values indicating more curvature. The intensity at $\mu$=1/2 from the quadratic law is,
\begin{align}
I(1/2)=1-c_1(1/2)-c_2(1/2)^2
\end{align}
and from our definition of $\delta=I(1/2)-I_{\rm lin}(1/2)$,
\begin{align}
\delta=\frac{1-\ell-c_1}{2}-\frac{c_2}{4}
\end{align}
so,
\begin{align}
c_1&=1-\ell-4\delta\\
c_2&=4\delta
\end{align}
and the intensity $I$($\mu$) is then,
\begin{align}
I(\mu) &= \ell + (1 - \ell + 4\delta)\mu - 4\delta\mu^2.
\end{align}
This parameterization is more intuitive than existing quadratic limb-darkening laws, and can be used to better understand how deviant a fit quadratic limb darkening law is from a linear function. The parameters $\ell$ and $\delta$ are uniquely determined by the intensity profile so are non-degenerate if the data quality is sufficient. The $(1 - \ell + 4\delta)$ term captures the correlation between parameters, which is linear. 

The re-parameterization is not designed to be used as the variables fit in a transit light curve model, as there are other less-correlated quadratic re-parameterizations available (see. \citealt{2013MNRAS.435.2152K}) and a conversion can easily be done after an optimized fit. Given the $(q_1,q_2)$ parameterization was found to have a Lucy-Sweeny type bias when limb-darkening is weak
\citep{2024AJ....168..227C}, we also advocate fitting with ($u_+$,$u_-$) or ($c_1$,$c_2$) with sufficiently wide uninformative priors.
Given the correlation between $\ell$ and $\delta$ is linear, affine invariant samplers such as \texttt{emcee} \citep{2013PASP..125..306F} will effortlessly handle the correlation. When fitting for the $u_+$ and $u_-$ quadratic parameterization, defined as
\begin{align}
u_+=c_1+c_2\\
u_-=c_1-c_2
\end{align}
the conversion becomes
\begin{align}
\ell &= 1-u_+, \label{eq:l_up}\\
\delta &= \frac{u_+-u_-}{8}. \label{eq:l_um}
\end{align}
Alternatively, for $q_1$ and $q_2$,
\begin{align}
\ell&=1-\sqrt{q_1},\\
\delta&=\frac{\sqrt{q_1}}{4}(1-2q_2).
\end{align}


\section{Results} \label{sec:results}

\subsection{Limb Darkening Data-Model Comparisons}

For each of the JWST transit light curves, we compared the empirically derived quadratic LD coefficients to the stellar atmospheric model values. We limited our analysis to only the quadratic LD law, for several reasons: (1) model selection statistics found the quadratic law to be optimal for most of our stars and (2) the JWST light curves were able to constrain both coefficients simultaneously, while higher-order laws suffered from unconstrained (or poorly constrained) coefficients. 

\begin{figure}[t!]
\centering
\includegraphics[width=\linewidth]{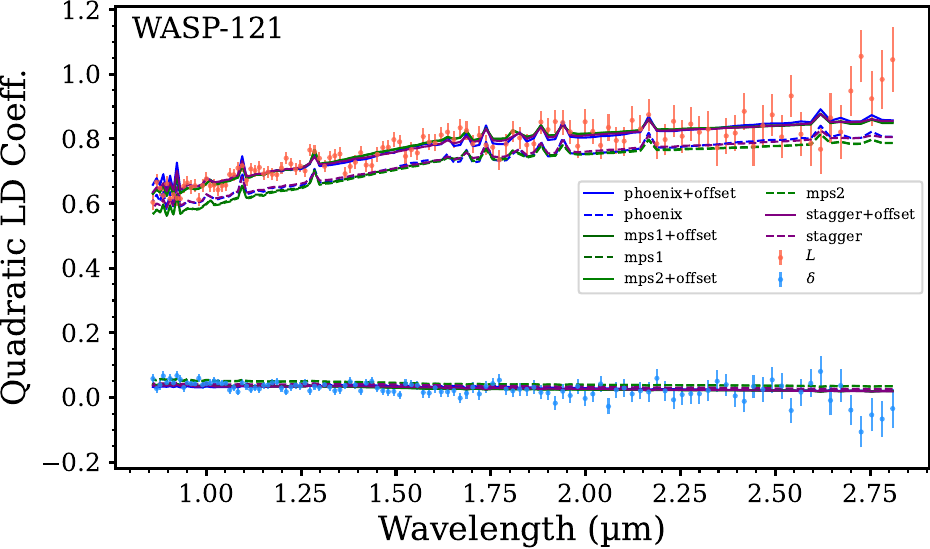}
\vspace{0.5em}
\includegraphics[width=\linewidth]{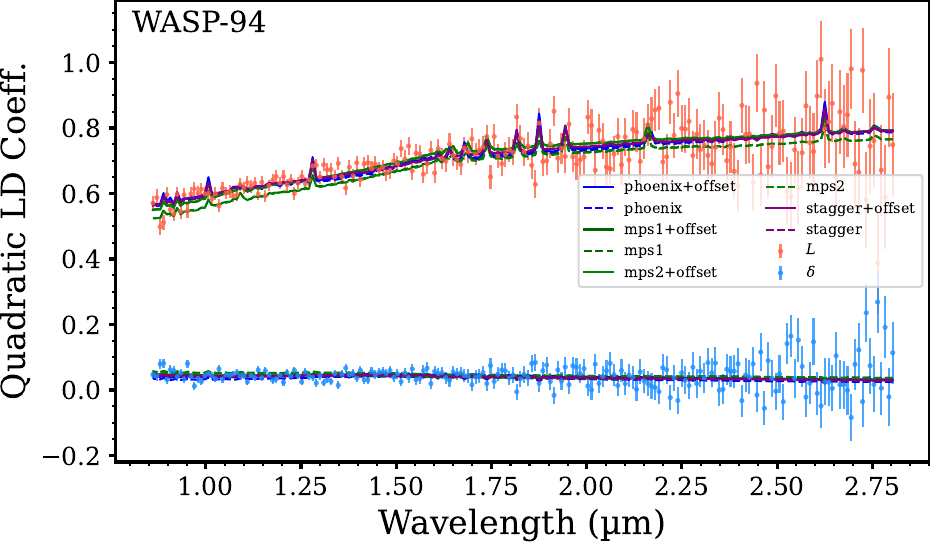}
\vspace{0.5em}
\includegraphics[width=\linewidth]{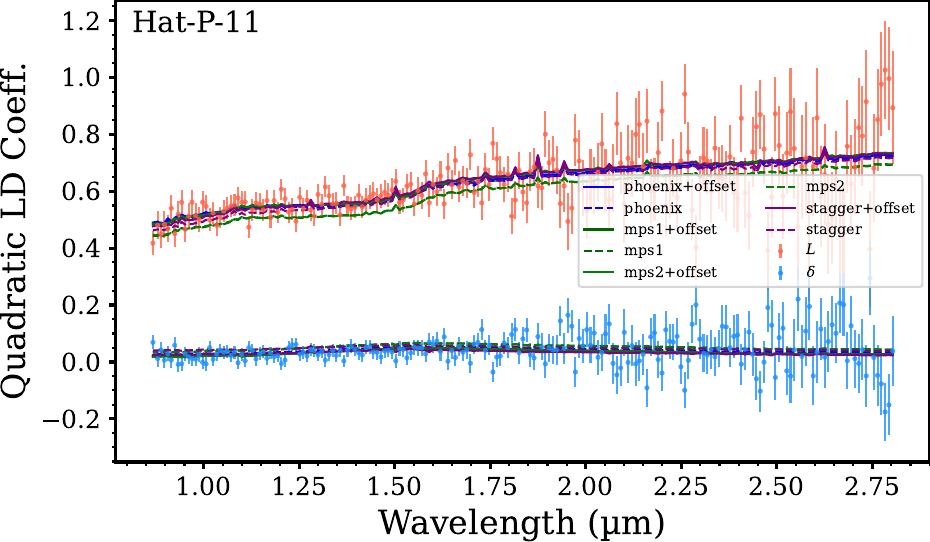}
\vspace{-0.5cm}
\caption{JWST-fit quadratic limb-darkening coefficient spectra for WASP-121, WASP-94, and HAT-P-11, compared with stellar atmosphere model predictions using $\mu_{\rm min} = 0.2$. The empirically derived coefficients $\ell$ (limb intensity) and $\delta$ (curvature) are shown with $1\sigma$ uncertainties. Model predictions from PHOENIX, MPS-ATLAS (sets 1 and 2), and Stagger are shown both with (solid) and without (dashed) the recommended offsets applied.}
\label{fig:ld_quad_examples_mu02}
\end{figure}

\begin{figure}[t!]
\centering
\includegraphics[width=\linewidth]{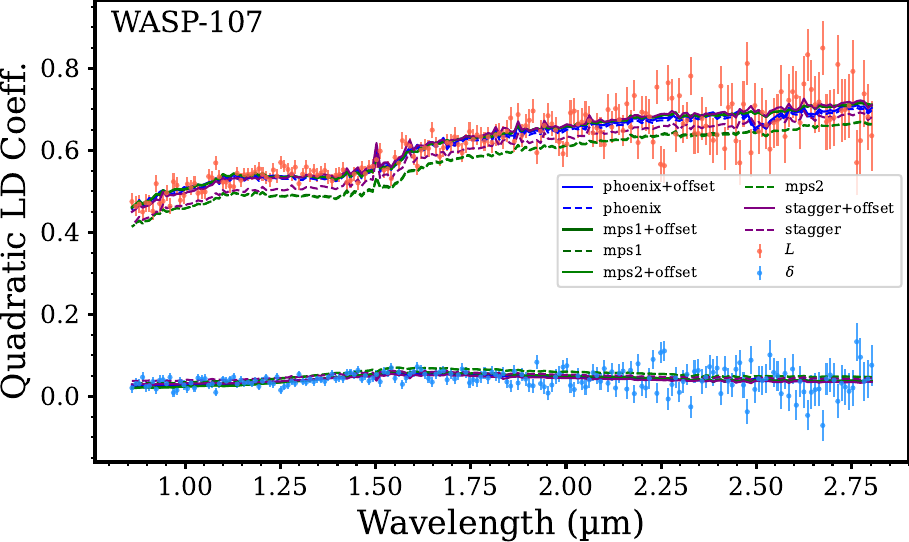}
\vspace{0.5em}
\includegraphics[width=\linewidth]{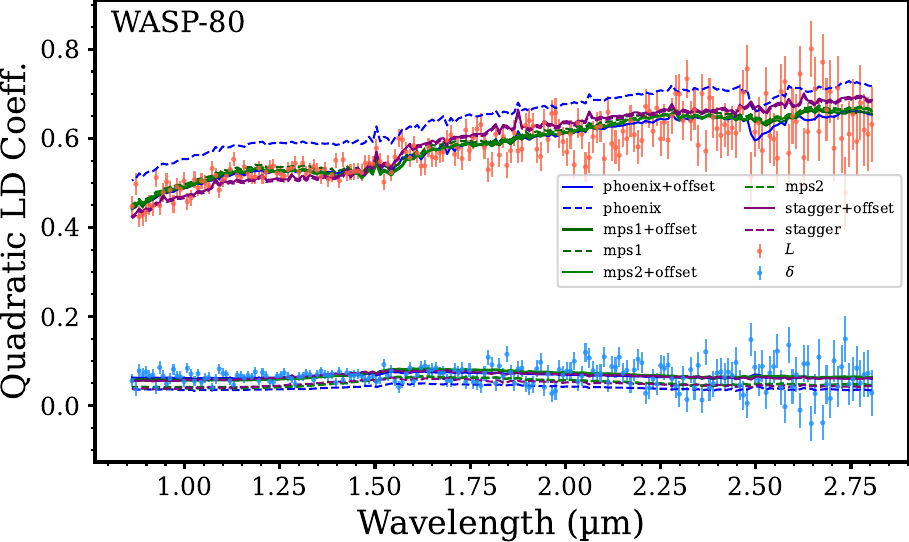}
\includegraphics[width=\linewidth]{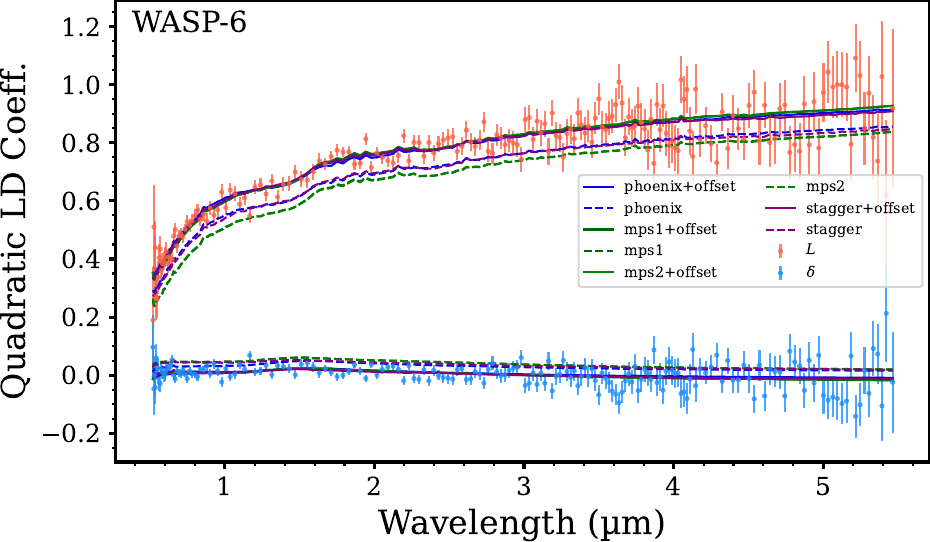}
\vspace{-0.5cm}
\caption{Same as Fig.~\ref{fig:ld_quad_examples_mu02}, but for WASP-107, WASP-80, and WASP-6.}
\label{fig:ld_quad_examples2_mu02}
\end{figure}

The JWST light curves were fit using the quadratic parameters $(u_+,u_-)$. For interpretation, these coefficients were then transformed to $(\ell,\delta)$ for each MCMC sample.
The resulting LD-coefficient spectra can be seen in Figs. \ref{fig:ld_quad_examples_mu02}  \& \ref{fig:ld_quad_examples2_mu02}.

When comparing the stellar models to the fit LD-coefficient spectra, we noticed for each FGK stellar type the models  accurately reproduce the wavelength-to-wavelength difference and overall shape of the coefficient spectra, but often have wavelength independent `gray' offsets in both
coefficients $(\ell,\delta)$. These offsets have also been observed in JWST prism observations of WASP-39~b \citep{Rustamkulov2022ApJ...928L...7R} and NIRSpec G395H observations of WASP-107~b \citep{2024Natur.630..831S}. 

\begin{figure}[t!]
\centering
\includegraphics[width=0.95\linewidth]{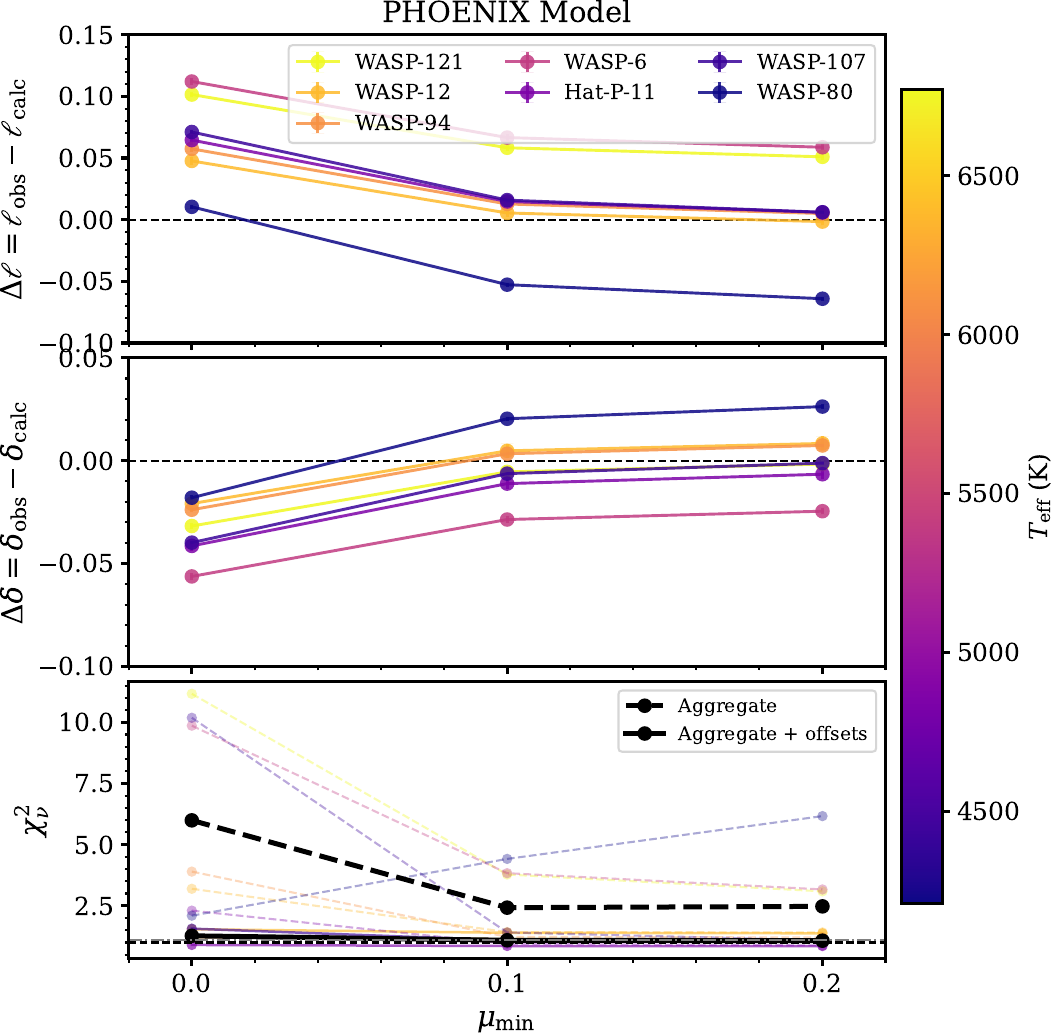}
\vspace{-0.25cm}
\caption{PHOENIX model limb-darkening comparison as a function of $\mu_{\rm min}$. \textit{Top:} Offset in the linear coefficient, $\Delta\ell$, for each target. \textit{Middle:} Offset in the curvature coefficient, $\Delta\delta$. \textit{Bottom:} Reduced chi-square ($\chi^2_\nu$) for each target (colored lines) and the aggregate statistics summing across all systems (black), shown both with and without offsets applied. Points are colored by stellar effective temperature.}
\label{fig:model_offsets_phoenix}
\end{figure}

\begin{figure}[t!]
\centering
\includegraphics[width=0.95\linewidth]{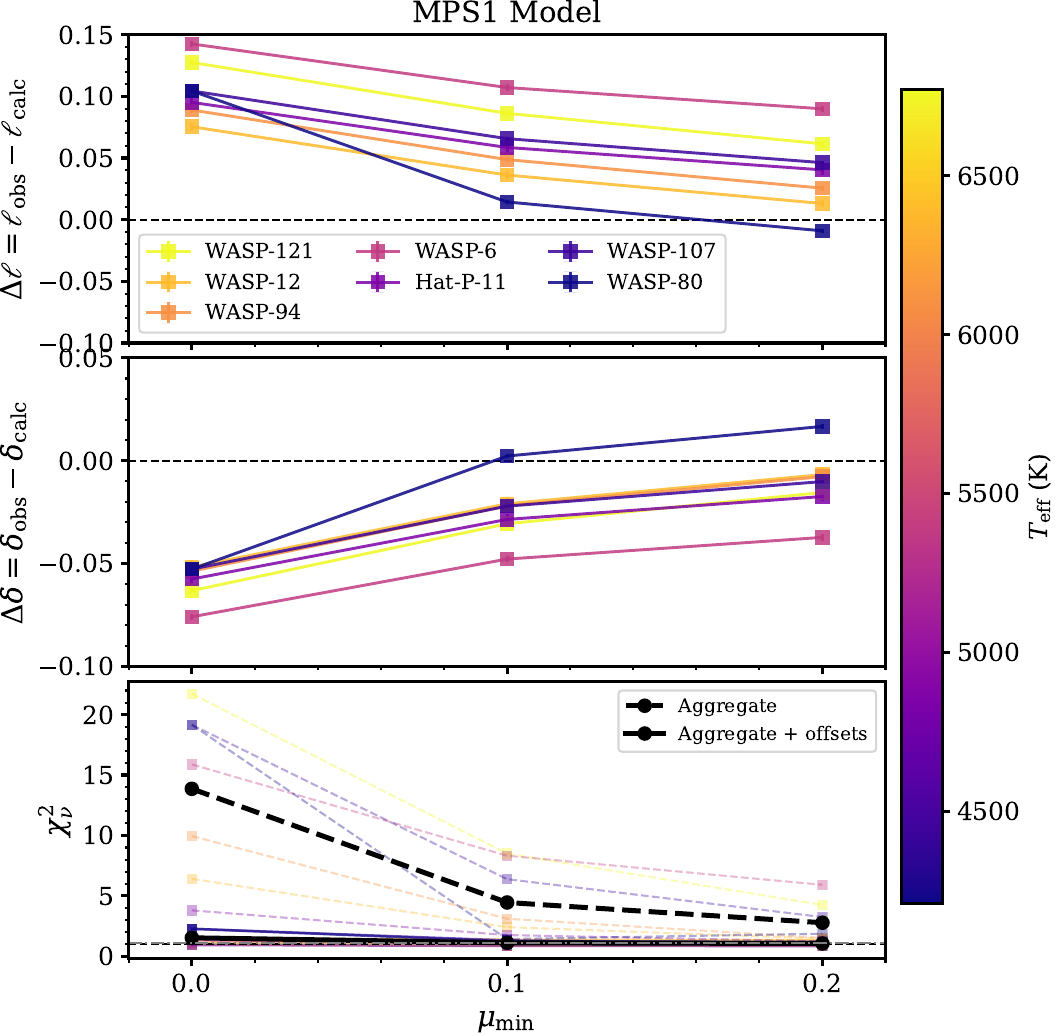}
\vspace{-0.25cm}
\caption{Same as Fig. \ref{fig:model_offsets_phoenix} but for the MPS1 models.}
\label{fig:model_offsets_mps1}
\end{figure}

\begin{figure}[t!]
\centering
\includegraphics[width=0.95\linewidth]{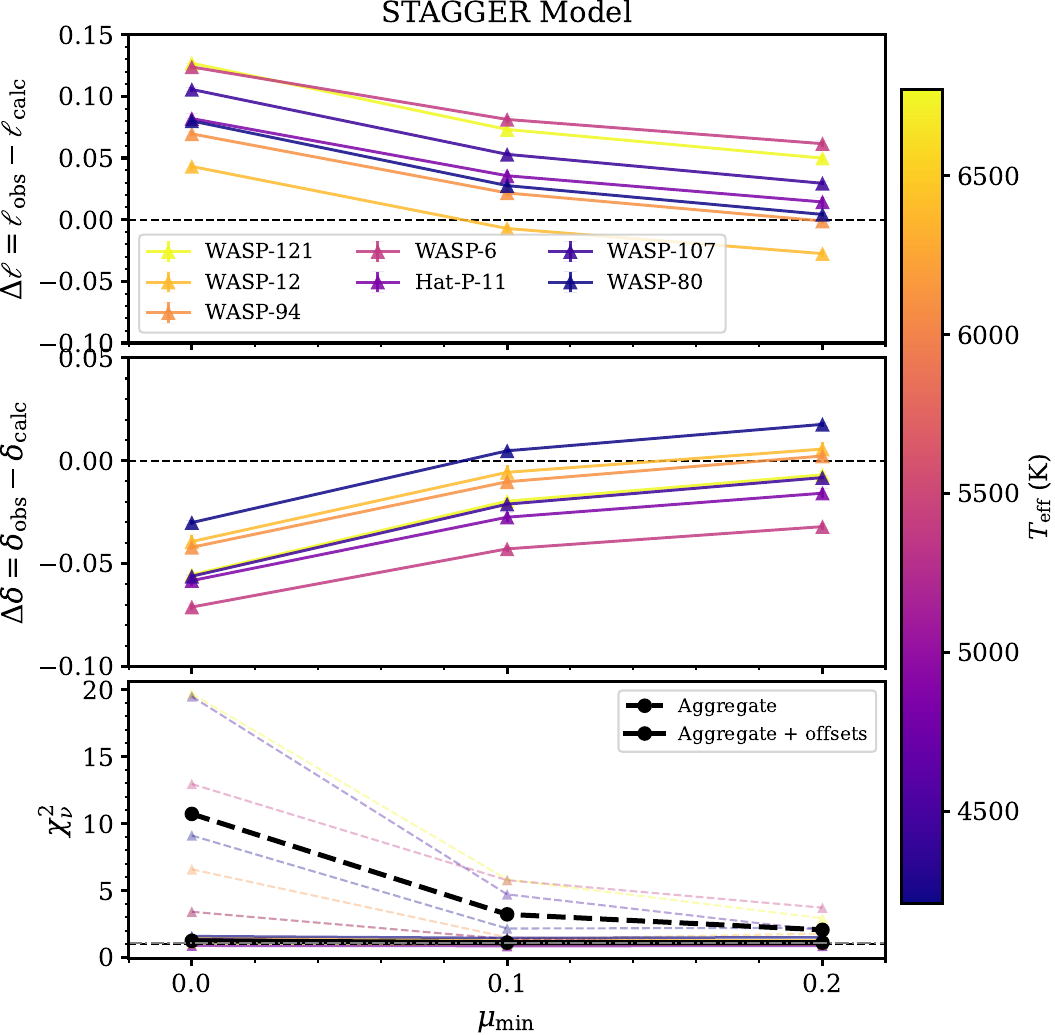}
\vspace{-0.25cm}
\caption{Same as Fig. \ref{fig:model_offsets_phoenix} but for the Stagger models.}
\label{fig:model_offsets_stagger}
\end{figure}

For each atmospheric model and JWST dataset, we calculated the $\chi^2$ goodness of fit relative to the LD-coefficient spectral data. We also calculated $\ell$ \& $\delta$ model offsets $(\Delta\ell,\Delta\delta)$ using the weighted average difference between the data and model. These offsets were calculated for each of our targets and at various values of $\mu_{\rm min}$ including 0, 0.1, \& 0.2. For the PHOENIX models, we used the numerical method to rescale $\mu$. The results can be seen in Figs. \ref{fig:model_offsets_phoenix}, \ref{fig:model_offsets_mps1}, \& \ref{fig:model_offsets_stagger}. 
We find that these offsets primarily depend on the stellar model used, as well as on the value of $\mu_{\rm min}$ used in the coefficient calculation. 
  
The models tend to overpredict the overall limb-darkening strength, as quantified by the linear coefficient $\ell$.
In general, this over-prediction is larger for 1-D models compared to 3-D models which predict weaker overall LD. This 1-D/3-D difference is well understood \citep{2012A&A...539A.102H}, as the 3-D hydrodynamic models have shallower temperature gradients compared to 1-D models. 
The measured stellar LD coefficients consistently indicate weaker overall limb darkening, i.e., higher $\ell$ values, across our FGK-star sample. This is consistent with hydrodynamic 3D simulations more accurately capturing the stellar convection and temperature-pressure profiles of these stars.

The offset is also larger and $\Delta\ell$ higher for lower values of $\mu_{\rm min}$. This trend can be understood numerically, as including more intensity points closer to the limb in the coefficient calculation tends to lower $\ell$. As model $\ell$-values tend to be too low compared to observations, using higher values of $\mu_{\rm min}$ helps compensate for this effect, leading to better overall fits. As seen in Figs. \ref{fig:model_offsets_phoenix}, \ref{fig:model_offsets_mps1}, \& \ref{fig:model_offsets_stagger}, both 1-D spherical and PP as well as the 3-D produce better fits to the data (better $\chi^2$) with $\mu_{\rm min}$=0.2, regardless of the inclusion of fit offsets.

Grey model offsets are seen in the LD curvature parameter $\delta$ as well when comparing against the fit data. In general, all targets have values of $\delta$ that fit near zero and are generally $<0.1$. This indicates that the fit stellar LD does not deviate strongly from a linear function, which helps to explain why the quadratic law performs well empirically. However, models generally predict more curvature resulting in slightly larger predicted values for $\delta$ than observed, unless the limbs are discarded from the coefficient calculation with a $\mu_{\rm min}$ of 0.2 minimizing the difference. One may suspect that a near-linear fit for the LD profile may be due to a lack of flexibility for the quadratic law. However, when we re-construct the posterior intensity profiles for the 3-parameter and 4-parameter LD laws, they match extremely well with the quadratic intensity profile posterior, deviating by $<0.5\%$ at $\mu=1/2$ (see Fig. \ref{fig:w6_i}). Thus, both higher-order LD laws including the quadratic contains sufficient flexibility to empirically fit the stellar LD. 
Fig. \ref{fig:w6_i} also illustrates how higher-order laws struggle to constrain the outermost limb intensities ($\mu<0.1$), even with high-quality JWST data. However, this outermost region represents $<1\%$ of the stellar disk area and even less of the emitted flux.
We note a similar behavior to Fig. \ref{fig:w6_i} can be seen in \cite{2023MNRAS.519.3723M} when recovering 4-parameter law LD coefficients for simulated Kepler transit light curves.

These LD model offsets cannot be explained by uncertainties in stellar $T_{\rm eff}$. To test this, we recalculated the LD coefficients for each stellar model by varying $T_{\rm eff}$ by a typical measurement uncertainty of $\pm50$ K with $\mu_{\rm min}=0.2$. We found that $\ell$ changed by only $\sim$0.0033, and $\delta$ by only $\sim$0.00025. These variations are approximately an order of magnitude smaller than the observed offsets (Table \ref{tab:offset_models}).

\subsection{The Effects of Magnetic fields on Limb Darkening}
Magnetic fields can modify stellar limb darkening and may help explain offsets between observed LD coefficients and models
\citep{2023MNRAS.519.3723M,2024NatAs...8..929K}. To investigate the effects of magnetic fields, we calculated limb-darkening coefficients using the MURaM 3D radiative magnetohydrodynamic simulations  
\citep{kostogryz2026effectsurfacemagneticfields,2005A&A...429..335V,2021A&A...653A..65W}. We used the K4 stellar model which has magnetic fields available up to 300 G\footnote{available at https://doi.org/10.17617/3.FBTIYY}, and compared the LD to our results. This comparison was done for WASP-6, which has a $\log[R'_{\rm HK}]= -4.741$ (\citealt{sing2016_nature}; see Table \ref{tab:obs}) making it the most active star in the current study where the classical activity measurement is defined \citep{1984ApJ...279..763N}.  
For each of the available magnetic stellar models, we calculated quadratic limb-darkening coefficients using the intensities with $\mu$ angles of 0.2 and larger. We then computed $\ell$ and $\delta$, and compared them to the JWST-fit coefficients. We note that WASP-6 has a stellar $T_{\rm eff}$ that is about 100 K hotter than the available K4 model.

\begin{figure}[t!]
\centering
\includegraphics[width=\linewidth]{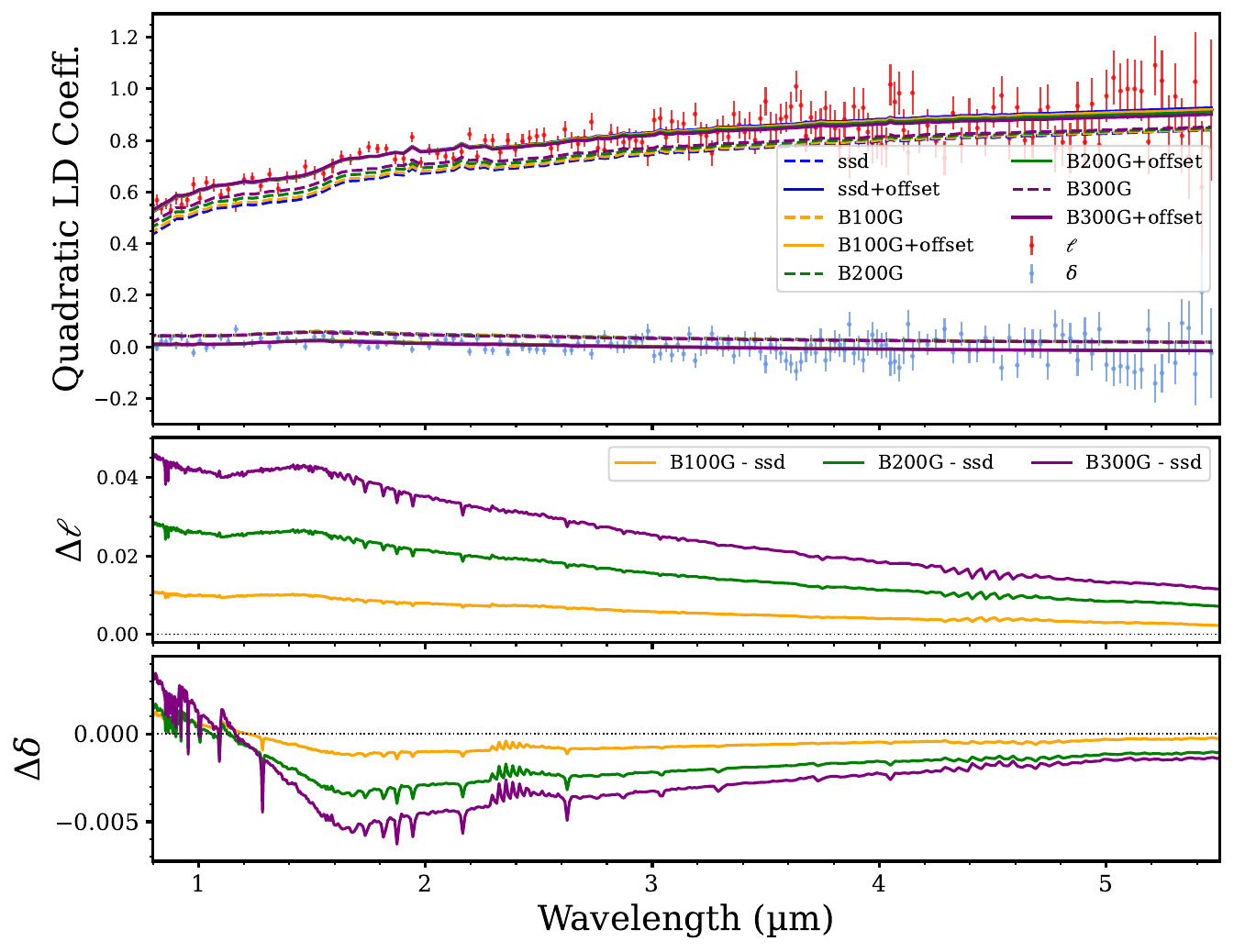}
\vspace{-0.6cm}
\caption{(Top) Same as Fig.~\ref{fig:ld_quad_examples_mu02}, but for WASP-6 using the MURaM magnetic models: with small-scale turbulent dynamo (SSD) baseline model (blue) and B-fields of 100 G, 200 G, and 300 G (orange, greed, purple respectively) shown. The change in LD parameters $\ell$ (Middle) and $\delta$ (bottom) between the non-magnetic model and model with B-fields is also shown.}
\label{fig:bmodel}
\end{figure}

We find that the introduction of magnetic fields alters the coefficients largely in the direction of the observed offsets, with the overall effect of weakening the predicted strength of limb-darkening, especially at the limb (see Fig. \ref{fig:bmodel}).   
Introducing magnetic fields to the MURaM 3D models increases $\ell$ by up to 0.04 at 1.4 $\mu$m for 300 G and 0.01 for 100 G. As shown in \cite{kostogryz2026effectsurfacemagneticfields}, there is also a sizable wavelength-dependence, with the effects of the magnetic fields weakening in the near-IR, with the 300 G model increasing $\ell$ by only 0.015 at 5 $\mu$m.
The effect on $\delta$ is smaller, with $\delta$ decreasing by -0.005 at 300 G and -0.001 for 100 G at 1.4 $\mu$m.  Thus, the overall effect of adding magnetic fields is to weaken the limb darkening at the limb, alter the intensity-profile to be more linear in nature, and introduce a wavelength-dependence.
The magnitude of the change in $\ell$ with magnetic fields can be large compared to the typical uncertainties in $T_{\rm eff}$. For example, for the Phoenix or MPS1 model of WASP-6, changing the $T_{\rm eff}$ by 100 K only changes $\ell$ by about 0.0066. 

We compared the MURaM models to the JWST-fit limb-darkening coefficients of WASP-6, fitting for constant offsets to the magnetic models (see Fig. \ref{fig:bmodel}). We find that the 200 G MURaM model provides the best fit among the magnetic models for WASP-6, with a $\chi^2$ only $\Delta\chi^2=4.01$ larger than the non-magnetic Stagger model.
Calculating the AIC \citep{2026ApJS..283...10T}, we find this corresponds to 20:1 odds in favor of the non-magnetic stagger model vs the 200 G MURaM model. Thus, by these metrics, the observed JWST limb-darkening is still a slightly better-fit non-magnetic stellar models. However, the general improvement within the MURaM model framework when introducing magnetic fields is notable.  

Introducing magnetic fields into the models tends to decrease the strength of limb-darkening \citep{kostogryz2026effectsurfacemagneticfields}, which goes in the general direction of the model offsets seen in Table \ref{tab:offset_models}. For WASP-6, the non-magnetic 3D Stagger model requires offsets of $\Delta\ell=+0.0616\pm 0.0025$ and $\Delta\delta=-0.0321\pm0.0016$ to match the JWST-derived coefficients. Given that the 300 G MURaM model increases $\ell$ by up to 0.04, approximately two-thirds of the Stagger model's $\ell$ offset could potentially be accounted for by the introduction of magnetic fields.
These findings suggest that magnetic fields could play an important role in explaining the differences between JWST-derived coefficients and the systematic offsets seen in stellar models. Future studies comparing larger samples across a range of stellar activity levels with denser grids of 3D magnetohydrodynamic models could help elucidate the effects of magnetic fields on limb darkening. Positive magnetic field signatures in LD studies should manifest as smaller derived coefficient offsets and improved fits to the wavelength-dependent LD coefficients. For WASP-6, incorporating magnetic fields into the MURaM models did reduce the required offset in $\ell$; however, the magnetic models did not provide superior fits to the chromatic LD coefficients compared to the non-magnetic Stagger models.

\subsection{Recommended LD models vs. stellar type and coefficient prior ranges}
To assess the overall performance of a given stellar model, we calculated the aggregate reduced $\chi_\nu^2$, defined as the sum of the individual $\chi^2$ values over all planets for a given $\mu$ divided by the total degrees of freedom. Results as a function of $\mu_{\rm min}$ are shown in Figs.~\ref{fig:model_offsets_phoenix}, \ref{fig:model_offsets_mps1}, and \ref{fig:model_offsets_stagger} for models with and without offsets $(\Delta\ell,\Delta\delta)$. This comparison leads to several conclusions. First, all of the stellar models produce better intensity fits when $\mu_{\rm min}=0.2$; we therefore recommend adopting $\mu_{\rm min}=0.2$ in most cases. In addition, including offsets generally improves the fits, such that acceptable fits to the transit data can typically be achieved ($\chi_\nu^2 \sim 1$). When fitting JWST transit light curves, we generally recommend first freely fitting the limb-darkening coefficients to measure the model offsets, before re-fitting with priors based on the offset-corrected models. Assuming the offset models reproduce the fit coefficients, this hybrid approach can help recover the planet transmission spectrum while reducing biases compared to only relying on stellar models.

In Table \ref{tab:offset_models}, we estimated the coefficient offsets between the stellar LD models and the FGK transit data, with $\mu_{\rm min}$ values of 0.1 and 0.2 reported. For each model, we report the mean observed-calculated offsets for both coefficients, $(\Delta\ell,\Delta\delta)$, along with their transformed quadratic counterparts ($u_+$,$u_-$) and ($c_1$,$c_2$). We also report on the empirical star-to-star scatter ($\sigma_{\Delta\ell}$, $\sigma_{\Delta\delta}$), which can be interpreted as the 1-$\sigma$ prior width on the offsets. These offset values and prior widths can be used for transit analyses where the limb-darkening is constrained using priors based on stellar atmosphere model values. $(\Delta\ell$ and $\Delta\delta)$ can help account for the model offsets observed in this study, while the prior ranges can help inform how deviant a particular star may be from these offsets. 

\begin{figure}[ht!]
\centering
\includegraphics[width=0.95\linewidth]{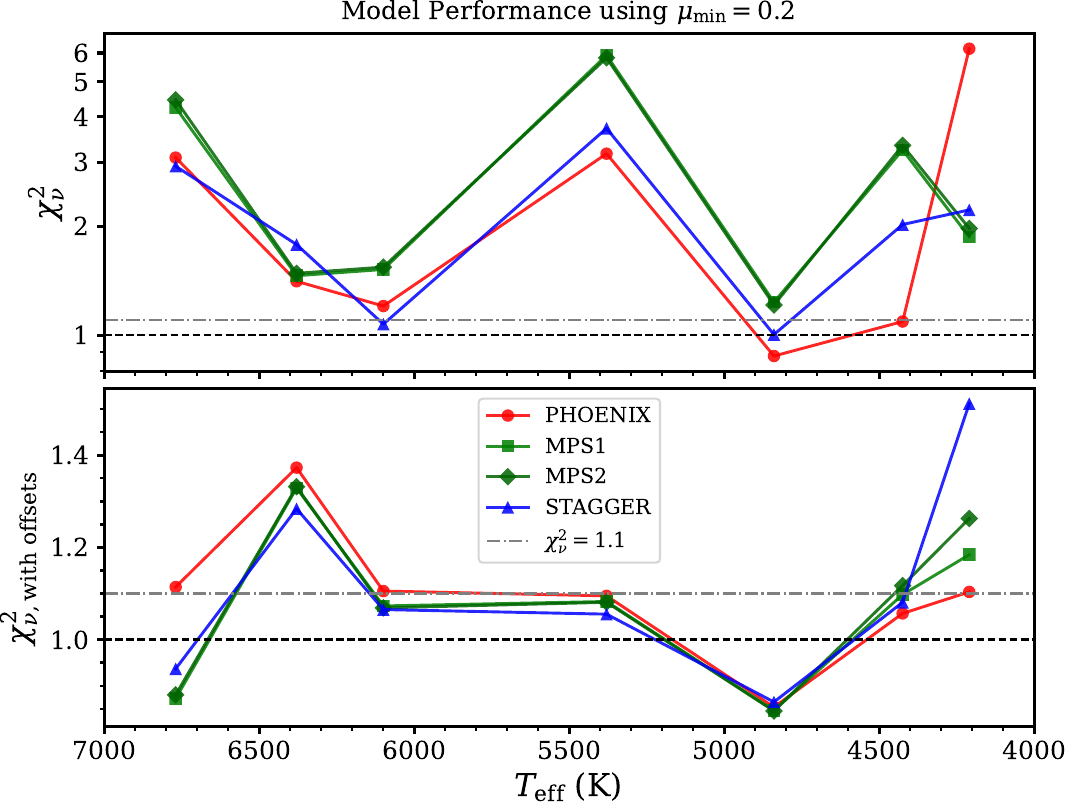}
\vspace{-0.1cm}
\caption{Model performance comparison as a function of stellar effective temperature, using $\mu_{\rm min} = 0.2$. \textit{Top:}  $\chi^2_\nu$ without offsets applied, showing the raw model performance. \textit{Bottom:} $\chi^2_\nu$ with gray offsets applied, demonstrating that all models can achieve acceptable fits ($\chi^2_\nu \lesssim 1.1$, dot-dashed lines) when offsets are included. Note the top panel is has a different y-range and is plotted on a log scale compared to the bottom panel on a linear scale.}
\label{fig:offset_chi2}
\end{figure}

Typical values of $\ell$ in the JWST/SOSS order 1 wavelength range span $\sim0.4$--0.8. Thus, offsets of $\Delta\ell \lesssim 0.036$ correspond to fractional biases of $\lesssim 7\%$ for all models. As shown in Fig.~\ref{fig:offset_chi2}, some model--star combinations, such as the PHOENIX and Stagger grids for HAT-P-11A ($T_{\rm eff}=4840$ K), yield statistically acceptable fits to the data without offsets ($\chi^2_\nu \lesssim 1.1$). Other stars require offsets before acceptable fits can be achieved. We estimate an expected star-to-star scatter when adopting model LD with offset corrections in Table \ref{tab:offset_models} using the empirical dispersion about the central values, $\sigma_{\Delta\ell}$ and $\sigma_{\Delta\delta}$. The observed scatter implies a typical fractional precision of $\sim 5\%$–10\% in $\ell$. This estimate will improve with additional JWST observations; however, for the current sample of $N=7$ systems, the sample standard deviation remains uncertain at the $\sim30\%$ level.
Figure~\ref{fig:offset_chi2} shows the $\chi_\nu^2$ values relative to the best-fitting stellar model as a function of $T_{\rm eff}$. 


\begin{table*}[ht!]
\centering
\caption{Recommended limb-darkening coefficient-offset priors for each stellar-atmosphere model at $\mu_{\rm min}=0.1$ and $0.2$. Offsets are defined as observed minus calculated. The quoted uncertainty is the empirical star-to-star scatter and is intended as the $1\sigma$ prior width for an individual system, not the uncertainty on the mean.}
\begin{tabular}{lccccccc}
\toprule
Model & $N$ & $\Delta \ell$ & $\Delta \delta$ & $\Delta u_+$ & $\Delta u_-$ & $\Delta c_1$ & $\Delta c_2$ \\
\midrule
\multicolumn{8}{c}{$\mu_{\rm min}=0.1$} \\
\midrule
PHOENIX & 7 & $+0.009 \pm 0.040$ & $-0.001 \pm 0.015$ & $-0.009 \pm 0.040$ & $-0.005 \pm 0.091$ & $-0.008 \pm 0.033$ & $-0.002 \pm 0.062$ \\
MPS1 & 7 & $+0.055 \pm 0.031$ & $-0.021 \pm 0.015$ & $-0.055 \pm 0.031$ & $+0.107 \pm 0.096$ & $+0.025 \pm 0.036$ & $-0.082 \pm 0.062$ \\
MPS2 & 7 & $+0.057 \pm 0.030$ & $-0.019 \pm 0.016$ & $-0.057 \pm 0.030$ & $+0.095 \pm 0.099$ & $+0.018 \pm 0.038$ & $-0.077 \pm 0.063$ \\
STAGGER & 7 & $+0.042 \pm 0.031$ & $-0.015 \pm 0.016$ & $-0.042 \pm 0.031$ & $+0.077 \pm 0.107$ & $+0.016 \pm 0.047$ & $-0.061 \pm 0.063$ \\
\midrule
\multicolumn{8}{c}{$\mu_{\rm min}=0.2$} \\
\midrule
PHOENIX & 7 & $~~0.000 \pm 0.041$ & $+0.004 \pm 0.016$ & $~~0.000 \pm 0.041$ & $-0.034 \pm 0.093$ & $-0.017 \pm 0.034$ & $+0.017 \pm 0.064$ \\
MPS1 & 7 & $+0.034 \pm 0.033$ & $-0.007 \pm 0.017$ & $-0.034 \pm 0.033$ & $+0.023 \pm 0.103$ & $-0.006 \pm 0.038$ & $-0.029 \pm 0.066$ \\
MPS2 & 7 & $+0.036 \pm 0.031$ & $-0.006 \pm 0.017$ & $-0.036 \pm 0.031$ & $+0.013 \pm 0.105$ & $-0.012 \pm 0.040$ & $-0.025 \pm 0.066$ \\
STAGGER & 7 & $+0.020 \pm 0.031$ & $-0.003 \pm 0.016$ & $-0.020 \pm 0.031$ & $-0.001 \pm 0.111$ & $-0.012 \pm 0.049$ & $-0.011 \pm 0.065$ \\
\bottomrule
\end{tabular}
\label{tab:offset_models}
\end{table*}

\section{Conclusion} \label{sec:concl}
We have presented a study of stellar limb-darkening using JWST transit observations for seven exoplanets  spanning FGK stellar types. The wide wavelength coverage from the red-optical to the near-infrared, combined with high signal-to-noise data, enabled us to place strong empirical constraints on the wavelength dependence of stellar limb darkening. We also presented a new quadratic limb-darkening transformation to better physically interpret the coefficients.
From this analysis, we draw the following key conclusions regarding the limb-darkening parameterizations used to fit JWST light curves and the performance of stellar atmosphere models:
\begin{itemize}
    \item The quadratic limb-darkening law is statistically preferred for JWST transit fitting and introduces minimal bias in the derived transit depth.
    \item Empirical FGK stellar limb darkening is more linear than models predict, with curvature parameters $\delta \lesssim 0.1$, and models also tend to over-predict the strength of limb darkening.
    \item All stellar atmosphere models produce better fits to the data when adopting $\mu_{\rm min} = 0.2$ for the limb-intensity calculation.
    \item The offsets in limb-darkening coefficients between the models and data are gray and small ($<7\%$ in $\ell$) using empirically-optimized limb-intensity calculations; we provide recommended offset priors for the PHOENIX, MPS-ATLAS, and Stagger grids.
\end{itemize}
We also revised methods for calculating limb-darkening coefficients using the spherical PHOENIX models, motivating the location of the stellar limb in the grid physically. For JWST spectroscopic transit analyses, we recommend first fitting for gray LD offsets, adopting a $\mu_{\rm min} = 0.2$ for stellar model LD coefficients, then re-fitting setting priors on the coefficients with offsets applied. Notably, excluding intensities at $\mu < 0.2$ affects only $\sim$4\% of the stellar disk area at the outer edges, yet significantly improves the agreement between models and data.
With optimized limb-intensity calculations using offsets and a $\mu_{\rm min} = 0.2$, the star-to-star scatter implies a typical fractional precision of $\sim10\%$ in $\ell$, which characterizes the overall strength of limb-darkening.

This study can be expanded by improving the statistics of FGK stars with additional observations and further investigating stellar models which include magnetic fields. In addition, similar studies can be done with transiting A and M-dwarfs to see if the models perform similarly well with the coefficient calculations methods used here, including the optimized use of $\mu_{\rm min}=0.2$ and $\mu_{\text{crit}}$ derived with the $\tau=1$ method. Studies comparing hybrid synthetic-photometry/atmosphere-model (SPAM) LDCs \citep{Howarth2011} to JWST derived empirical values would also be beneficial.

As stellar limb darkening directly probes a star's temperature gradient and opacity structure, these JWST transit data offer new empirical benchmarks that can inform improvements to stellar atmosphere models across a wide range of spectral types. Such model improvements can directly feedback and improve the transmission spectra measured for exoplanets.


\section*{ACKNOWLEDGEMENTS}
We thank the anonymous referee for their constructive comments. This work is based on observations made with the NASA/ESA/CSA James Webb Space Telescope. The data were obtained from the Mikulski Archive for Space Telescopes at the Space Telescope Science Institute, which is operated by the Association of Universities for Research in Astronomy, Inc., under NASA contract NAS 5-03127 for JWST. These observations are associated with programs \#1201 \& 5924. Support for JWST program GO-5924 was provided by NASA through a grant from the Space Telescope Science Institute, which is operated by the Association of Universities for Research in Astronomy, Inc., under NASA contract NAS 5-26555. 
This work has made use of data from the European Space Agency (ESA) mission
{\it Gaia} (\url{https://www.cosmos.esa.int/gaia}), processed by the {\it Gaia}
Data Processing and Analysis Consortium (DPAC,
\url{https://www.cosmos.esa.int/web/gaia/dpac/consortium}). Funding for the DPAC
has been provided by national institutions, in particular the institutions
participating in the {\it Gaia} Multilateral Agreement.

\software{NumPy \citep{numpy}, SciPy \citep{scipy}, MatPlotLib \citep{matplotlib}, AstroPy \citep{astropy}, Exotic-LD \citep{2024JOSS....9.6816G}}
\facilities{ JWST(NIRSpec, NIRISS) }
\section*{Data Availability} 
All of the data presented in this article were obtained from the Mikulski Archive for Space Telescopes (MAST) at the Space Telescope Science Institute. The specific observations analyzed can be accessed via \dataset[doi:10.17909/hd77-3s11]{https://doi.org/10.17909/hd77-3s11}.
\bibliography{main}{}

@ARTICLE{2021A&A...653A..65W,
       author = {{Witzke}, V. and {Shapiro}, A.~I. and {Cernetic}, M. and {Tagirov}, R.~V. and {Kostogryz}, N.~M. and {Anusha}, L.~S. and {Unruh}, Y.~C. and {Solanki}, S.~K. and {Kurucz}, R.~L.},
        title = "{MPS-ATLAS: A fast all-in-one code for synthesising stellar spectra}",
      journal = {\aap},
         year = 2021,
        month = sep,
       volume = {653},
          eid = {A65},
        pages = {A65},
          doi = {10.1051/0004-6361/202140275},
archivePrefix = {arXiv},
       eprint = {2105.13611},
 primaryClass = {astro-ph.SR},
       adsurl = {https://ui.adsabs.harvard.edu/abs/2021A&A...653A..65W}
}

@ARTICLE{2005A&A...429..335V,
       author = {{V{\"o}gler}, A. and {Shelyag}, S. and {Sch{\"u}ssler}, M. and {Cattaneo}, F. and {Emonet}, T. and {Linde}, T.},
        title = "{Simulations of magneto-convection in the solar photosphere.  Equations, methods, and results of the MURaM code}",
      journal = {\aap},
         year = 2005,
        month = jan,
       volume = {429},
        pages = {335-351},
          doi = {10.1051/0004-6361:20041507},
       adsurl = {https://ui.adsabs.harvard.edu/abs/2005A&A...429..335V}
}

@ARTICLE{2022MNRAS.510.4857A,
       author = {{Ahrer}, E. and {Wheatley}, P.~J. and {Kirk}, J. and {Gandhi}, S. and {King}, G.~W. and {Louden}, T.},
        title = "{LRG-BEASTS: Sodium absorption and Rayleigh scattering in the atmosphere of WASP-94A b using NTT/EFOSC2}",
      journal = {\mnras},
         year = 2022,
        month = mar,
       volume = {510},
       number = {4},
        pages = {4857-4871},
          doi = {10.1093/mnras/stab3805},
archivePrefix = {arXiv},
       eprint = {2201.02212},
 primaryClass = {astro-ph.EP},
       adsurl = {https://ui.adsabs.harvard.edu/abs/2022MNRAS.510.4857A}
}

@ARTICLE{Basilicata2024,
       author = {{Basilicata}, M. and {Giacobbe}, P. and {Bonomo}, A.~S. and {Scandariato}, G. and {Brogi}, M. and {Singh}, V. and {Di Paola}, A. and {Mancini}, L. and {Sozzetti}, A. and {Lanza}, A.~F. and {Cubillos}, P.~E. and {Damasso}, M. and {Desidera}, S. and {Biazzo}, K. and {Bignamini}, A. and {Borsa}, F. and {Cabona}, L. and {Carleo}, I. and {Ghedina}, A. and {Guilluy}, G. and {Maggio}, A. and {Mainella}, G. and {Micela}, G. and {Molinari}, E. and {Molinaro}, M. and {Nardiello}, D. and {Pedani}, M. and {Pino}, L. and {Poretti}, E. and {Southworth}, J. and {Stangret}, M. and {Turrini}, D.},
        title = "{The GAPS Programme at TNG. LV. Multiple molecular species in the atmosphere of HAT-P-11 b and review of the HAT-P-11 planetary system}",
      journal = {\aap},
         year = 2024,
        month = jun,
       volume = {686},
          eid = {A127},
        pages = {A127},
          doi = {10.1051/0004-6361/202347659},
archivePrefix = {arXiv},
       eprint = {2403.01527},
 primaryClass = {astro-ph.EP},
       adsurl = {https://ui.adsabs.harvard.edu/abs/2024A&A...686A.127B}
}

@article{Howarth2011,
    author = {Howarth, Ian D.},
    title = {On stellar limb darkening and exoplanetary transits},
    journal = {Monthly Notices of the Royal Astronomical Society},
    volume = {418},
    number = {2},
    pages = {1165-1175},
    year = {2011},
    month = {12},
    issn = {0035-8711},
    doi = {10.1111/j.1365-2966.2011.19568.x},
    url = {https://doi.org/10.1111/j.1365-2966.2011.19568.x},
    eprint = {https://academic.oup.com/mnras/article-pdf/418/2/1165/3707266/mnras0418-1165.pdf},
}

@ARTICLE{1984ApJ...279..763N,
       author = {{Noyes}, R.~W. and {Hartmann}, L.~W. and {Baliunas}, S.~L. and {Duncan}, D.~K. and {Vaughan}, A.~H.},
        title = "{Rotation, convection, and magnetic activity in lower main-sequence stars.}",
      journal = {\apj},
         year = 1984,
        month = apr,
       volume = {279},
        pages = {763-777},
          doi = {10.1086/161945},
       adsurl = {https://ui.adsabs.harvard.edu/abs/1984ApJ...279..763N}
}

@ARTICLE{2024NatAs...8..929K,
       author = {{Kostogryz}, Nadiia M. and {Shapiro}, Alexander I. and {Witzke}, Veronika and {Cameron}, Robert H. and {Gizon}, Laurent and {Krivova}, Natalie A. and {Ludwig}, Hans-G. and {Maxted}, Pierre F.~L. and {Seager}, Sara and {Solanki}, Sami K. and {Valenti}, Jeff},
        title = "{Magnetic origin of the discrepancy between stellar limb-darkening models and observations}",
      journal = {Nature Astronomy},
         year = 2024,
        month = jul,
       volume = {8},
        pages = {929-937},
          doi = {10.1038/s41550-024-02252-5},
       adsurl = {https://ui.adsabs.harvard.edu/abs/2024NatAs...8..929K}
}

@misc{kostogryz2026effectsurfacemagneticfields,
      title={Effect of surface magnetic fields on limb darkening in main-sequence stars}, 
      author={N. Kostogryz and A. I. Shapiro and V. Witzke and T. Bhatia and S. K. Solanki and I. Kuhlemann and V. Vasilyev and Y. C. Unruh},
      year={2026},
      eprint={2606.21912},
      archivePrefix={arXiv},
      primaryClass={astro-ph.SR},
      url={https://arxiv.org/abs/2606.21912}, 
}

@ARTICLE{2017AJ....154..111M,
       author = {{Morello}, G. and {Tsiaras}, A. and {Howarth}, I.~D. and {Homeier}, D.},
        title = "{High-precision Stellar Limb-darkening in Exoplanetary Transits}",
      journal = {\aj},
         year = 2017,
        month = sep,
       volume = {154},
       number = {3},
          eid = {111},
        pages = {111},
          doi = {10.3847/1538-3881/aa8405},
archivePrefix = {arXiv},
       eprint = {1704.08232},
 primaryClass = {astro-ph.EP},
       adsurl = {https://ui.adsabs.harvard.edu/abs/2017AJ....154..111M}
}

@ARTICLE{2024ApJ...977L...7K,
       author = {{Keers}, Rosa E. and {Shapiro}, Alexander I. and {Kostogryz}, Nadiia M. and {Glidden}, Ana and {Niraula}, Prajwal and {Rackham}, Benjamin V. and {Seager}, Sara and {Solanki}, Sami K. and {Unruh}, Yvonne C. and {Vasilyev}, Valeriy and {de Wit}, Julien},
        title = "{Reliable Transmission Spectrum Extraction with a Three-parameter Limb-darkening Law}",
      journal = {\apjl},
         year = 2024,
        month = dec,
       volume = {977},
       number = {1},
          eid = {L7},
        pages = {L7},
          doi = {10.3847/2041-8213/ad8b51},
archivePrefix = {arXiv},
       eprint = {2410.18617},
 primaryClass = {astro-ph.EP},
       adsurl = {https://ui.adsabs.harvard.edu/abs/2024ApJ...977L...7K}
}

@ARTICLE{2026NatAs..10..258K,
       author = {{Krishnamurthy}, Vigneshwaran and {Carteret}, Yann and {Piaulet-Ghorayeb}, Caroline and {Splinter}, Jared and {Doshi}, Dhvani and {Radica}, Michael and {Coulombe}, Louis-Philippe and {Allart}, Romain and {Bourrier}, Vincent and {Cowan}, Nicolas B. and {Doyon}, Ren{\'e} and {Lafreni{\`e}re}, David and {Albert}, Lo{\"\i}c and {Benneke}, Bj{\"o}rn and {Dang}, Lisa and {Jayawardhana}, Ray and {Johnstone}, Doug and {Kaltenegger}, Lisa and {Langeveld}, Adam B. and {Pelletier}, Stefan and {Rowe}, Jason F. and {Roy}, Pierre-Alexis and {Taylor}, Jake and {Turner}, Jake D.},
        title = "{Continuous helium absorption from both the leading and trailing tails of WASP-107 b}",
      journal = {Nature Astronomy},
         year = 2026,
        month = feb,
       volume = {10},
        pages = {258-270},
          doi = {10.1038/s41550-025-02710-8},
archivePrefix = {arXiv},
       eprint = {2505.20588},
 primaryClass = {astro-ph.EP},
       adsurl = {https://ui.adsabs.harvard.edu/abs/2026NatAs..10..258K}
}

@ARTICLE{2026A&A...706A...2P,
       author = {{Pelletier}, S. and {Coulombe}, L.-P. and {Splinter}, J. and {Benneke}, B. and {MacDonald}, R.~J. and {Lafreni{\`e}re}, D. and {Cowan}, N.~B. and {Allart}, R. and {Rauscher}, E. and {Frazier}, R.~C. and {Meyer}, M.~R. and {Albert}, L. and {Dang}, L. and {Doyon}, R. and {Ehrenreich}, D. and {Flagg}, L. and {Johnstone}, D. and {Langeveld}, A.~B. and {Lim}, O. and {Piaulet-Ghorayeb}, C. and {Radica}, M. and {Rowe}, J. and {Taylor}, J. and {Turner}, J.~D.},
        title = "{Enriched volatiles and refractories but deficient titanium on the day-side atmosphere of WASP-121b revealed by JWST/NIRISS}",
      journal = {\aap},
         year = 2026,
        month = jan,
       volume = {706},
          eid = {A2},
        pages = {A2},
          doi = {10.1051/0004-6361/202556985},
archivePrefix = {arXiv},
       eprint = {2508.18341},
 primaryClass = {astro-ph.EP},
       adsurl = {https://ui.adsabs.harvard.edu/abs/2026A&A...706A...2P}
}

@ARTICLE{2025NatCo..1610822A,
       author = {{Allart}, Romain and {Coulombe}, Louis-Philippe and {Carteret}, Yann and {Splinter}, Jared and {Dang}, Lisa and {Bourrier}, Vincent and {Lafreni{\`e}re}, David and {Albert}, Lo{\"\i}c and {Artigau}, {\'E}tienne and {Benneke}, Bj{\"o}rn and {Cowan}, Nicolas B. and {Doyon}, Ren{\'e} and {Krishnamurthy}, Vigneshwaran and {Jayawardhana}, Ray and {Johnstone}, Doug and {Langeveld}, Adam B. and {Meyer}, Michael R. and {Pelletier}, Stefan and {Piaulet-Ghorayeb}, Caroline and {Radica}, Michael and {Taylor}, Jake and {Turner}, Jake D.},
        title = "{A complex structure of escaping helium spanning more than half the orbit of the ultra-hot Jupiter WASP-121 b}",
      journal = {Nature Communications},
         year = 2025,
        month = dec,
       volume = {16},
       number = {1},
          eid = {10822},
        pages = {10822},
          doi = {10.1038/s41467-025-66628-5},
archivePrefix = {arXiv},
       eprint = {2510.09809},
 primaryClass = {astro-ph.EP},
       adsurl = {https://ui.adsabs.harvard.edu/abs/2025NatCo..1610822A}
}

@ARTICLE{2015MNRAS.453.3821P,
       author = {{Parviainen}, H. and {Aigrain}, S.},
        title = "{LDTK: Limb Darkening Toolkit}",
      journal = {\mnras},
         year = 2015,
        month = nov,
       volume = {453},
       number = {4},
        pages = {3821-3826},
          doi = {10.1093/mnras/stv1857},
archivePrefix = {arXiv},
       eprint = {1508.02634},
 primaryClass = {astro-ph.EP},
       adsurl = {https://ui.adsabs.harvard.edu/abs/2015MNRAS.453.3821P}
}

@ARTICLE{2015MNRAS.450.1879E,
       author = {{Espinoza}, N{\'e}stor and {Jord{\'a}n}, Andr{\'e}s},
        title = "{Limb darkening and exoplanets: testing stellar model atmospheres and identifying biases in transit parameters}",
      journal = {\mnras},
         year = 2015,
        month = jun,
       volume = {450},
       number = {2},
        pages = {1879-1899},
          doi = {10.1093/mnras/stv744},
archivePrefix = {arXiv},
       eprint = {1503.07020},
 primaryClass = {astro-ph.EP},
       adsurl = {https://ui.adsabs.harvard.edu/abs/2015MNRAS.450.1879E}
}

@ARTICLE{2026ApJS..283...10T,
       author = {{Thorngren}, Daniel P. and {Sing}, David K. and {Mukherjee}, Sagnick},
        title = "{Bayesian Model Comparison and Significance: Widespread Errors and How to Correct Them}",
      journal = {\apjs},
         year = 2026,
        month = mar,
       volume = {283},
       number = {1},
          eid = {10},
        pages = {10},
          doi = {10.3847/1538-4365/ae0e71},
archivePrefix = {arXiv},
       eprint = {2510.00169},
 primaryClass = {astro-ph.EP},
       adsurl = {https://ui.adsabs.harvard.edu/abs/2026ApJS..283...10T}
}

@ARTICLE{2024AJ....168..227C,
       author = {{Coulombe}, Louis-Philippe and {Roy}, Pierre-Alexis and {Benneke}, Bj{\"o}rn},
        title = "{Biases in Exoplanet Transmission Spectra Introduced by Limb-darkening Parametrization}",
      journal = {\aj},
         year = 2024,
        month = nov,
       volume = {168},
       number = {5},
          eid = {227},
        pages = {227},
          doi = {10.3847/1538-3881/ad7aef},
archivePrefix = {arXiv},
       eprint = {2409.03812},
 primaryClass = {astro-ph.EP},
       adsurl = {https://ui.adsabs.harvard.edu/abs/2024AJ....168..227C}
}

@ARTICLE{2018A&A...618A..20C,
       author = {{Claret}, Antonio},
        title = "{A new method to compute limb-darkening coefficients for stellar atmosphere models with spherical symmetry: the space missions TESS, Kepler, CoRoT, and MOST}",
      journal = {\aap},
         year = 2018,
        month = oct,
       volume = {618},
          eid = {A20},
        pages = {A20},
          doi = {10.1051/0004-6361/201833060},
archivePrefix = {arXiv},
       eprint = {1804.10135},
 primaryClass = {astro-ph.SR},
       adsurl = {https://ui.adsabs.harvard.edu/abs/2018A&A...618A..20C}
}

@ARTICLE{2008A&A...491..633L,
       author = {{Lester}, J.~B. and {Neilson}, H.~R.},
        title = "{satlas: spherical versions of the atlas stellar atmosphere program}",
      journal = {\aap},
         year = 2008,
        month = nov,
       volume = {491},
       number = {2},
        pages = {633-641},
          doi = {10.1051/0004-6361:200810578},
archivePrefix = {arXiv},
       eprint = {0809.1870},
 primaryClass = {astro-ph},
       adsurl = {https://ui.adsabs.harvard.edu/abs/2008A&A...491..633L}
}

@ARTICLE{2013A&A...554A..98N,
       author = {{Neilson}, H.~R. and {Lester}, J.~B.},
        title = "{Spherically-symmetric model stellar atmospheres and limb darkening. I. Limb-darkening laws, gravity-darkening coefficients and angular diameter corrections for red giant stars}",
      journal = {\aap},
         year = 2013,
        month = jun,
       volume = {554},
          eid = {A98},
        pages = {A98},
          doi = {10.1051/0004-6361/201321502},
archivePrefix = {arXiv},
       eprint = {1305.1311},
 primaryClass = {astro-ph.SR},
       adsurl = {https://ui.adsabs.harvard.edu/abs/2013A&A...554A..98N}
}

@ARTICLE{2013A&A...556A..86N,
       author = {{Neilson}, H.~R. and {Lester}, J.~B.},
        title = "{Spherically symmetric model stellar atmospheres and limb darkening. II. Limb-darkening laws, gravity-darkening coefficients and angular diameter corrections for FGK dwarf stars}",
      journal = {\aap},
         year = 2013,
        month = aug,
       volume = {556},
          eid = {A86},
        pages = {A86},
          doi = {10.1051/0004-6361/201321888},
archivePrefix = {arXiv},
       eprint = {1306.6640},
 primaryClass = {astro-ph.SR},
       adsurl = {https://ui.adsabs.harvard.edu/abs/2013A&A...556A..86N}
}

@ARTICLE{2003A&A...412..241C,
       author = {{Claret}, A. and {Hauschildt}, P.~H.},
        title = "{The limb-darkening for spherically symmetric NextGen model atmospheres: A-G main-sequence and sub-giant stars}",
      journal = {\aap},
         year = 2003,
        month = dec,
       volume = {412},
        pages = {241-248},
          doi = {10.1051/0004-6361:20031405},
       adsurl = {https://ui.adsabs.harvard.edu/abs/2003A&A...412..241C}
}

@article{husser2013new,
  title={A new extensive library of PHOENIX stellar atmospheres and synthetic spectra},
  author={Husser, T-O and Wende-von Berg, Sebastian and Dreizler, Stefan and Homeier, Derek and Reiners, Ansgar and Barman, Travis and Hauschildt, Peter H},
  journal={Astronomy \& Astrophysics},
  volume={553},
  year={2013},
  publisher={EDP Sciences}
}

@article{magic2015stagger,
  title={The Stagger-grid: A grid of 3D stellar atmosphere models-IV. Limb darkening coefficients},
  author={Magic, Zazralt and Chiavassa, Andrea and Collet, Remo and Asplund, Martin},
  journal={Astronomy \& Astrophysics},
  volume={573},
  pages={A90},
  year={2015},
  publisher={EDP Sciences}
}

@article{kostogryz2023mps,
  title={MPS-ATLAS library of stellar model atmospheres and spectra},
  author={Kostogryz, N and Shapiro, AI and Witzke, V and Grant, D and Wakeford, HR and Stevenson, KB and Solanki, SK and Gizon, L},
  journal={Research Notes of the AAS},
  volume={7},
  number={3},
  pages={39},
  year={2023},
  publisher={The American Astronomical Society}
}

@article{kostogryz2022stellar,
  title={Stellar limb darkening. A new MPS-ATLAS library for Kepler, TESS, CHEOPS, and PLATO passbands},
  author={Kostogryz, NM and Witzke, V and Shapiro, AI and Solanki, SK and Maxted, PFL and Kurucz, RL and Gizon, L},
  journal={Astronomy \& Astrophysics},
  volume={666},
  pages={A60},
  year={2022},
  publisher={EDP Sciences}
}

@ARTICLE{2024AJ....168..231S,
       author = {{Sing}, David K. and {Evans-Soma}, Thomas M. and {Rustamkulov}, Zafar and {Lothringer}, Joshua D. and {Mayne}, Nathan J. and {Schlaufman}, Kevin C.},
        title = "{An Absolute Mass, Precise Age, and Hints of Planetary Winds for WASP-121A and b from a JWST NIRSpec Phase Curve}",
      journal = {\aj},
         year = 2024,
        month = dec,
       volume = {168},
       number = {6},
          eid = {231},
        pages = {231},
          doi = {10.3847/1538-3881/ad7fe7},
archivePrefix = {arXiv},
       eprint = {2501.03844},
 primaryClass = {astro-ph.EP},
       adsurl = {https://ui.adsabs.harvard.edu/abs/2024AJ....168..231S}
}

@ARTICLE{2013MNRAS.435.2152K,
       author = {{Kipping}, David M.},
        title = "{Efficient, uninformative sampling of limb darkening coefficients for two-parameter laws}",
      journal = {\mnras},
         year = 2013,
        month = nov,
       volume = {435},
       number = {3},
        pages = {2152-2160},
          doi = {10.1093/mnras/stt1435},
archivePrefix = {arXiv},
       eprint = {1308.0009},
 primaryClass = {astro-ph.SR},
       adsurl = {https://ui.adsabs.harvard.edu/abs/2013MNRAS.435.2152K}
}

@ARTICLE{2010A&A...510A..21S,
       author = {{Sing}, D.~K.},
        title = "{Stellar limb-darkening coefficients for CoRot and Kepler}",
      journal = {\aap},
         year = 2010,
        month = feb,
       volume = {510},
          eid = {A21},
        pages = {A21},
          doi = {10.1051/0004-6361/200913675},
archivePrefix = {arXiv},
       eprint = {0912.2274},
 primaryClass = {astro-ph.EP},
       adsurl = {https://ui.adsabs.harvard.edu/abs/2010A&A...510A..21S}
}

@ARTICLE{2025arXiv251116771W,
       author = {{Wang}, Le-Chris and {Rustamkulov}, Zafar and {Sing}, David K. and {Lothringer}, Joshua and {McCreery}, Patrick and {Thorngren}, Daniel and {Alam}, Munazza K.},
        title = "{A Comprehensive Analysis of the Panchromatic Transmission Spectrum of the Hot-Saturn WASP-96 b: Nondetection of Haze, Possible Sodium Limb Asymmetry, Stellar Characterization, and Formation History}",
      journal = {arXiv e-prints},
         year = 2025,
        month = nov,
          eid = {arXiv:2511.16771},
        pages = {arXiv:2511.16771},
          doi = {10.48550/arXiv.2511.16771},
archivePrefix = {arXiv},
       eprint = {2511.16771},
 primaryClass = {astro-ph.EP},
       adsurl = {https://ui.adsabs.harvard.edu/abs/2025arXiv251116771W}
}

@ARTICLE{2001ApJ...552..699B,
   author = {{Brown}, T.~M. and {Charbonneau}, D. and {Gilliland}, R.~L. and 
	{Noyes}, R.~W. and {Burrows}, A.},
    title = "{Hubble Space Telescope Time-Series Photometry of the Transiting Planet of HD 209458}",
  journal = {\apj},
   eprint = {arXiv:astro-ph/0101336},
     year = 2001,
    month = may,
   volume = 552,
    pages = {699-709},
      doi = {10.1086/320580},
   adsurl = {http://adsabs.harvard.edu/abs/2001ApJ...552..699B}
}

@ARTICLE{2017JGRE..122...53D,
   author = {{Deming}, L.~D. and {Seager}, S.},
    title = "{Illusion and reality in the atmospheres of exoplanets}",
  journal = {JGRE},
     year = 2017,
    month = jan,
   volume = 122,
    pages = {53-75},
      doi = {10.1002/2016JE005155},
   adsurl = {http://adsabs.harvard.edu/abs/2017JGRE..122...53D}
}

@ARTICLE{2013PASP..125..306F,
   author = {{Foreman-Mackey}, D. and {Hogg}, D.~W. and {Lang}, D. and {Goodman}, J.
	},
    title = "{emcee: The MCMC Hammer}",
  journal = {\pasp},
archivePrefix = "arXiv",
   eprint = {1202.3665},
 primaryClass = "astro-ph.IM",
     year = 2013,
    month = mar,
   volume = 125,
    pages = {306},
      doi = {10.1086/670067},
   adsurl = {http://adsabs.harvard.edu/abs/2013PASP..125..306F}
}

@ARTICLE{2012A&A...539A.102H,
   author = {{Hayek}, W. and {Sing}, D. and {Pont}, F. and {Asplund}, M.},
    title = "{Limb darkening laws for two exoplanet host stars derived from 3D stellar model atmospheres. Comparison with 1D models and HST light curve observations}",
  journal = {\aap},
archivePrefix = "arXiv",
   eprint = {1202.0548},
 primaryClass = "astro-ph.SR",
     year = 2012,
    month = mar,
   volume = 539,
      eid = {A102},
    pages = {A102},
      doi = {10.1051/0004-6361/201117868},
   adsurl = {http://adsabs.harvard.edu/abs/2012A%26A...539A.102H}
}

@ARTICLE{2000ApJ...529L..41H,
   author = {{Henry}, G.~W. and {Marcy}, G.~W. and {Butler}, R.~P. and {Vogt}, S.~S.
	},
    title = "{A Transiting ``51 Peg-like'' Planet}",
  journal = {\apjl},
     year = 2000,
    month = jan,
   volume = 529,
    pages = {L41-L44},
      doi = {10.1086/312458},
   adsurl = {http://adsabs.harvard.edu/abs/2000ApJ...529L..41H}
}

@ARTICLE{2015PASP..127.1161K,
   author = {{Kreidberg}, L.},
    title = "{batman: BAsic Transit Model cAlculatioN in Python}",
  journal = {\pasp},
archivePrefix = "arXiv",
   eprint = {1507.08285},
 primaryClass = "astro-ph.EP",
     year = 2015,
    month = nov,
   volume = 127,
    pages = {1161},
      doi = {10.1086/683602},
   adsurl = {http://adsabs.harvard.edu/abs/2015PASP..127.1161K}
}

@ARTICLE{2008ApJ...686..658S,
   author = {{Sing}, D.~K. and {Vidal-Madjar}, A. and {D{\'e}sert}, J.-M. and 
	{Lecavelier des Etangs}, A. and {Ballester}, G.},
    title = "{Hubble Space Telescope STIS Optical Transit Transmission Spectra of the Hot Jupiter HD 209458b}",
  journal = {\apj},
archivePrefix = "arXiv",
   eprint = {0802.3864},
     year = 2008,
    month = oct,
   volume = 686,
    pages = {658-666},
      doi = {10.1086/590075},
   adsurl = {http://adsabs.harvard.edu/abs/2008ApJ...686..658S}
}

@ARTICLE{2024NatAs...8.1562M,
       author = {{Murphy}, Matthew M. and {Beatty}, Thomas G. and {Schlawin}, Everett and {Bell}, Taylor J. and {Line}, Michael R. and {Greene}, Thomas P. and {Parmentier}, Vivien and {Rauscher}, Emily and {Welbanks}, Luis and {Fortney}, Jonathan J. and {Rieke}, Marcia},
        title = "{Evidence for morning-to-evening limb asymmetry on the cool low-density exoplanet WASP-107 b}",
      journal = {Nature Astronomy},
         year = 2024,
        month = dec,
       volume = {8},
        pages = {1562-1574},
          doi = {10.1038/s41550-024-02367-9},
archivePrefix = {arXiv},
       eprint = {2406.09863},
 primaryClass = {astro-ph.EP},
       adsurl = {https://ui.adsabs.harvard.edu/abs/2024NatAs...8.1562M}
}

@article{2025arXiv250510910M,
author = {Sagnick Mukherjee  and David K. Sing  and Guangwei Fu  and Kevin B. Stevenson  and Stephen P. Schmidt  and Harry Baskett  and Mei Ting Mak  and Patrick McCreery  and Natalie H. Allen  and Katherine A. Bennett  and Duncan A. Christie  and Carlos Gascón  and Jayesh Goyal  and Éric Hébrard  and Joshua D. Lothringer  and Mercedes López-Morales  and Jacob Lustig-Yaeger  and Erin M. May  and L. C. Mayorga  and Nathan Mayne  and Lakeisha M. Ramos Rosado  and Henrique Reggiani  and Zafar Rustamkulov  and Kevin C. Schlaufman  and Kristin S. Sotzen  and Daniel Thorngren  and Le-Chris Wang  and Maria Zamyatina },
title = {Cloudy mornings and clear evenings on a gas giant exoplanet},
journal = {Science},
volume = {392},
number = {6800},
pages = {858-862},
year = {2026},
doi = {10.1126/science.adx5903},
URL = {https://www.science.org/doi/abs/10.1126/science.adx5903},
eprint = {https://www.science.org/doi/pdf/10.1126/science.adx5903}}

@ARTICLE{2024Natur.630..831S,
       author = {{Sing}, David K. and {Rustamkulov}, Zafar and {Thorngren}, Daniel P. and {Barstow}, Joanna K. and {Tremblin}, Pascal and {Alves de Oliveira}, Catarina and {Beck}, Tracy L. and {Birkmann}, Stephan M. and {Challener}, Ryan C. and {Crouzet}, Nicolas and {Espinoza}, N{\'e}stor and {Ferruit}, Pierre and {Giardino}, Giovanna and {Gressier}, Am{\'e}lie and {Lee}, Elspeth K.~H. and {Lewis}, Nikole K. and {Maiolino}, Roberto and {Manjavacas}, Elena and {Rauscher}, Bernard J. and {Sirianni}, Marco and {Valenti}, Jeff A.},
        title = "{A warm Neptune's methane reveals core mass and vigorous atmospheric mixing}",
      journal = {\nat},
         year = 2024,
        month = jun,
       volume = {630},
       number = {8018},
        pages = {831-835},
          doi = {10.1038/s41586-024-07395-z},
archivePrefix = {arXiv},
       eprint = {2405.11027},
 primaryClass = {astro-ph.EP},
       adsurl = {https://ui.adsabs.harvard.edu/abs/2024Natur.630..831S}
}

@ARTICLE{2024JOSS....9.6816G,
       author = {{Grant}, David and {Wakeford}, Hannah},
        title = "{ExoTiC-LD: thirty seconds to stellar limb-darkening coefficients}",
      journal = {The Journal of Open Source Software},
         year = 2024,
        month = aug,
       volume = {9},
       number = {100},
          eid = {6816},
        pages = {6816},
          doi = {10.21105/joss.06816},
archivePrefix = {arXiv},
       eprint = {2408.10341},
 primaryClass = {astro-ph.IM},
       adsurl = {https://ui.adsabs.harvard.edu/abs/2024JOSS....9.6816G}
}

@ARTICLE{2011MNRAS.417.2166S,
       author = {{Southworth}, John},
        title = "{Homogeneous studies of transiting extrasolar planets - IV. Thirty systems with space-based light curves}",
      journal = {\mnras},
         year = 2011,
        month = nov,
       volume = {417},
       number = {3},
        pages = {2166-2196},
          doi = {10.1111/j.1365-2966.2011.19399.x},
archivePrefix = {arXiv},
       eprint = {1107.1235},
 primaryClass = {astro-ph.EP},
       adsurl = {https://ui.adsabs.harvard.edu/abs/2011MNRAS.417.2166S}
}

@ARTICLE{2023MNRAS.519.3723M,
       author = {{Maxted}, Pierre F.~L.},
        title = "{Limb darkening measurements from TESS and Kepler light curves of transiting exoplanets}",
      journal = {\mnras},
         year = 2023,
        month = mar,
       volume = {519},
       number = {3},
        pages = {3723-3735},
          doi = {10.1093/mnras/stac3741},
archivePrefix = {arXiv},
       eprint = {2212.09117},
 primaryClass = {astro-ph.EP},
       adsurl = {https://ui.adsabs.harvard.edu/abs/2023MNRAS.519.3723M}
}

@ARTICLE{2015A&A...573A..90M,
       author = {{Magic}, Z. and {Chiavassa}, A. and {Collet}, R. and {Asplund}, M.},
        title = "{The Stagger-grid: A grid of 3D stellar atmosphere models. IV. Limb darkening coefficients}",
      journal = {\aap},
         year = 2015,
        month = jan,
       volume = {573},
          eid = {A90},
        pages = {A90},
          doi = {10.1051/0004-6361/201423804},
archivePrefix = {arXiv},
       eprint = {1403.3487},
 primaryClass = {astro-ph.SR},
       adsurl = {https://ui.adsabs.harvard.edu/abs/2015A&A...573A..90M}
}

@ARTICLE{cho16,
       author = {{Choi}, Jieun and {Dotter}, Aaron and {Conroy}, Charlie and
         {Cantiello}, Matteo and {Paxton}, Bill and {Johnson}, Benjamin D.},
        title = "{Mesa Isochrones and Stellar Tracks (MIST). I. Solar-scaled Models}",
      journal = {\apj},
         year = 2016,
        month = jun,
       volume = {823},
       number = {2},
          eid = {102},
        pages = {102},
          doi = {10.3847/0004-637X/823/2/102},
archivePrefix = {arXiv},
       eprint = {1604.08592},
 primaryClass = {astro-ph.SR},
       adsurl = {https://ui.adsabs.harvard.edu/abs/2016ApJ...823..102C}
}

@ARTICLE{dot16,
       author = {{Dotter}, Aaron},
        title = "{MESA Isochrones and Stellar Tracks (MIST) 0: Methods for the Construction of Stellar Isochrones}",
      journal = {\apjs},
         year = 2016,
        month = jan,
       volume = {222},
       number = {1},
          eid = {8},
        pages = {8},
          doi = {10.3847/0067-0049/222/1/8},
archivePrefix = {arXiv},
       eprint = {1601.05144},
 primaryClass = {astro-ph.SR},
       adsurl = {https://ui.adsabs.harvard.edu/abs/2016ApJS..222....8D}
}

@ARTICLE{gai16,
       author = {{Gaia Collaboration} and {Prusti}, T. and {de Bruijne}, J.~H.~J. and
         {Brown}, A.~G.~A. and {Vallenari}, A. and {Babusiaux}, C. and
         {Bailer-Jones}, C.~A.~L. and {Bastian}, U. and {Biermann}, M. and
         {Evans}, D.~W. and {Eyer}, L. and {Jansen}, F. and {Jordi}, C. and
         {Klioner}, S.~A. and {Lammers}, U. and {Lindegren}, L. and {Luri}, X. and
         {Mignard}, F. and {Milligan}, D.~J. and {Panem}, C. and
         {Poinsignon}, V. and {Pourbaix}, D. and {Randich}, S. and {Sarri}, G. and
         {Sartoretti}, P. and {Siddiqui}, H.~I. and {Soubiran}, C. and
         {Valette}, V. and {van Leeuwen}, F. and {Walton}, N.~A. and
         {Aerts}, C. and {Arenou}, F. and {Cropper}, M. and {Drimmel}, R. and
         {H{\o}g}, E. and {Katz}, D. and {Lattanzi}, M.~G. and {O'Mullane}, W. and
         {Grebel}, E.~K. and {Holland}, A.~D. and {Huc}, C. and {Passot}, X. and
         {Bramante}, L. and {Cacciari}, C. and {Casta{\~n}eda}, J. and
         {Chaoul}, L. and {Cheek}, N. and {De Angeli}, F. and {Fabricius}, C. and
         {Guerra}, R. and {Hern{\'a}ndez}, J. and {Jean-Antoine-Piccolo}, A. and
         {Masana}, E. and {Messineo}, R. and {Mowlavi}, N. and
         {Nienartowicz}, K. and {Ord{\'o}{\~n}ez-Blanco}, D. and {Panuzzo}, P. and
         {Portell}, J. and {Richards}, P.~J. and {Riello}, M. and
         {Seabroke}, G.~M. and {Tanga}, P. and {Th{\'e}venin}, F. and
         {Torra}, J. and {Els}, S.~G. and {Gracia-Abril}, G. and
         {Comoretto}, G. and {Garcia-Reinaldos}, M. and {Lock}, T. and
         {Mercier}, E. and {Altmann}, M. and {Andrae}, R. and
         {Astraatmadja}, T.~L. and {Bellas-Velidis}, I. and {Benson}, K. and
         {Berthier}, J. and {Blomme}, R. and {Busso}, G. and {Carry}, B. and
         {Cellino}, A. and {Clementini}, G. and {Cowell}, S. and {Creevey}, O. and
         {Cuypers}, J. and {Davidson}, M. and {De Ridder}, J. and
         {de Torres}, A. and {Delchambre}, L. and {Dell'Oro}, A. and
         {Ducourant}, C. and {Fr{\'e}mat}, Y. and {Garc{\'\i}a-Torres}, M. and
         {Gosset}, E. and {Halbwachs}, J. -L. and {Hambly}, N.~C. and
         {Harrison}, D.~L. and {Hauser}, M. and {Hestroffer}, D. and
         {Hodgkin}, S.~T. and {Huckle}, H.~E. and {Hutton}, A. and
         {Jasniewicz}, G. and {Jordan}, S. and {Kontizas}, M. and {Korn}, A.~J. and
         {Lanzafame}, A.~C. and {Manteiga}, M. and {Moitinho}, A. and
         {Muinonen}, K. and {Osinde}, J. and {Pancino}, E. and {Pauwels}, T. and
         {Petit}, J. -M. and {Recio-Blanco}, A. and {Robin}, A.~C. and
         {Sarro}, L.~M. and {Siopis}, C. and {Smith}, M. and {Smith}, K.~W. and
         {Sozzetti}, A. and {Thuillot}, W. and {van Reeven}, W. and {Viala}, Y. and
         {Abbas}, U. and {Abreu Aramburu}, A. and {Accart}, S. and
         {Aguado}, J.~J. and {Allan}, P.~M. and {Allasia}, W. and
         {Altavilla}, G. and {{\'A}lvarez}, M.~A. and {Alves}, J. and
         {Anderson}, R.~I. and {Andrei}, A.~H. and {Anglada Varela}, E. and
         {Antiche}, E. and {Antoja}, T. and {Ant{\'o}n}, S. and {Arcay}, B. and
         {Atzei}, A. and {Ayache}, L. and {Bach}, N. and {Baker}, S.~G. and
         {Balaguer-N{\'u}{\~n}ez}, L. and {Barache}, C. and {Barata}, C. and
         {Barbier}, A. and {Barblan}, F. and {Baroni}, M. and
         {Barrado y Navascu{\'e}s}, D. and {Barros}, M. and {Barstow}, M.~A. and
         {Becciani}, U. and {Bellazzini}, M. and {Bellei}, G. and
         {Bello Garc{\'\i}a}, A. and {Belokurov}, V. and {Bendjoya}, P. and
         {Berihuete}, A. and {Bianchi}, L. and {Bienaym{\'e}}, O. and
         {Billebaud}, F. and {Blagorodnova}, N. and {Blanco-Cuaresma}, S. and
         {Boch}, T. and {Bombrun}, A. and {Borrachero}, R. and {Bouquillon}, S. and
         {Bourda}, G. and {Bouy}, H. and {Bragaglia}, A. and {Breddels}, M.~A. and
         {Brouillet}, N. and {Br{\"u}semeister}, T. and {Bucciarelli}, B. and
         {Budnik}, F. and {Burgess}, P. and {Burgon}, R. and {Burlacu}, A. and
         {Busonero}, D. and {Buzzi}, R. and {Caffau}, E. and {Cambras}, J. and
         {Campbell}, H. and {Cancelliere}, R. and {Cantat-Gaudin}, T. and
         {Carlucci}, T. and {Carrasco}, J.~M. and {Castellani}, M. and
         {Charlot}, P. and {Charnas}, J. and {Charvet}, P. and {Chassat}, F. and
         {Chiavassa}, A. and {Clotet}, M. and {Cocozza}, G. and
         {Collins}, R.~S. and {Collins}, P. and {Costigan}, G. and {Crifo}, F. and
         {Cross}, N.~J.~G. and {Crosta}, M. and {Crowley}, C. and {Dafonte}, C. and
         {Damerdji}, Y. and {Dapergolas}, A. and {David}, P. and {David}, M. and
         {De Cat}, P. and {de Felice}, F. and {de Laverny}, P. and
         {De Luise}, F. and {De March}, R. and {de Martino}, D. and
         {de Souza}, R. and {Debosscher}, J. and {del Pozo}, E. and {Delbo}, M. and
         {Delgado}, A. and {Delgado}, H.~E. and {di Marco}, F. and
         {Di Matteo}, P. and {Diakite}, S. and {Distefano}, E. and
         {Dolding}, C. and {Dos Anjos}, S. and {Drazinos}, P. and
         {Dur{\'a}n}, J. and {Dzigan}, Y. and {Ecale}, E. and {Edvardsson}, B. and
         {Enke}, H. and {Erdmann}, M. and {Escolar}, D. and {Espina}, M. and
         {Evans}, N.~W. and {Eynard Bontemps}, G. and {Fabre}, C. and
         {Fabrizio}, M. and {Faigler}, S. and {Falc{\~a}o}, A.~J. and
         {Farr{\`a}s Casas}, M. and {Faye}, F. and {Federici}, L. and
         {Fedorets}, G. and {Fern{\'a}ndez-Hern{\'a}ndez}, J. and
         {Fernique}, P. and {Fienga}, A. and {Figueras}, F. and {Filippi}, F. and
         {Findeisen}, K. and {Fonti}, A. and {Fouesneau}, M. and {Fraile}, E. and
         {Fraser}, M. and {Fuchs}, J. and {Furnell}, R. and {Gai}, M. and
         {Galleti}, S. and {Galluccio}, L. and {Garabato}, D. and
         {Garc{\'\i}a-Sedano}, F. and {Gar{\'e}}, P. and {Garofalo}, A. and
         {Garralda}, N. and {Gavras}, P. and {Gerssen}, J. and {Geyer}, R. and
         {Gilmore}, G. and {Girona}, S. and {Giuffrida}, G. and {Gomes}, M. and
         {Gonz{\'a}lez-Marcos}, A. and {Gonz{\'a}lez-N{\'u}{\~n}ez}, J. and
         {Gonz{\'a}lez-Vidal}, J.~J. and {Granvik}, M. and {Guerrier}, A. and
         {Guillout}, P. and {Guiraud}, J. and {G{\'u}rpide}, A. and
         {Guti{\'e}rrez-S{\'a}nchez}, R. and {Guy}, L.~P. and {Haigron}, R. and
         {Hatzidimitriou}, D. and {Haywood}, M. and {Heiter}, U. and
         {Helmi}, A. and {Hobbs}, D. and {Hofmann}, W. and {Holl}, B. and {Holland
        }, G. and {Hunt}, J.~A.~S. and {Hypki}, A. and {Icardi}, V. and
         {Irwin}, M. and {Jevardat de Fombelle}, G. and {Jofr{\'e}}, P. and
         {Jonker}, P.~G. and {Jorissen}, A. and {Julbe}, F. and
         {Karampelas}, A. and {Kochoska}, A. and {Kohley}, R. and
         {Kolenberg}, K. and {Kontizas}, E. and {Koposov}, S.~E. and
         {Kordopatis}, G. and {Koubsky}, P. and {Kowalczyk}, A. and
         {Krone-Martins}, A. and {Kudryashova}, M. and {Kull}, I. and
         {Bachchan}, R.~K. and {Lacoste-Seris}, F. and {Lanza}, A.~F. and
         {Lavigne}, J. -B. and {Le Poncin-Lafitte}, C. and {Lebreton}, Y. and
         {Lebzelter}, T. and {Leccia}, S. and {Leclerc}, N. and
         {Lecoeur-Taibi}, I. and {Lemaitre}, V. and {Lenhardt}, H. and
         {Leroux}, F. and {Liao}, S. and {Licata}, E. and
         {Lindstr{\o}m}, H.~E.~P. and {Lister}, T.~A. and {Livanou}, E. and
         {Lobel}, A. and {L{\"o}ffler}, W. and {L{\'o}pez}, M. and
         {Lopez-Lozano}, A. and {Lorenz}, D. and {Loureiro}, T. and
         {MacDonald}, I. and {Magalh{\~a}es Fernandes}, T. and {Managau}, S. and
         {Mann}, R.~G. and {Mantelet}, G. and {Marchal}, O. and
         {Marchant}, J.~M. and {Marconi}, M. and {Marie}, J. and {Marinoni}, S. and
         {Marrese}, P.~M. and {Marschalk{\'o}}, G. and {Marshall}, D.~J. and
         {Mart{\'\i}n-Fleitas}, J.~M. and {Martino}, M. and {Mary}, N. and
         {Matijevi{\v{c}}}, G. and {Mazeh}, T. and {McMillan}, P.~J. and
         {Messina}, S. and {Mestre}, A. and {Michalik}, D. and {Millar}, N.~R. and
         {Miranda}, B.~M.~H. and {Molina}, D. and {Molinaro}, R. and
         {Molinaro}, M. and {Moln{\'a}r}, L. and {Moniez}, M. and
         {Montegriffo}, P. and {Monteiro}, D. and {Mor}, R. and {Mora}, A. and
         {Morbidelli}, R. and {Morel}, T. and {Morgenthaler}, S. and
         {Morley}, T. and {Morris}, D. and {Mulone}, A.~F. and {Muraveva}, T. and
         {Musella}, I. and {Narbonne}, J. and {Nelemans}, G. and {Nicastro}, L. and
         {Noval}, L. and {Ord{\'e}novic}, C. and {Ordieres-Mer{\'e}}, J. and
         {Osborne}, P. and {Pagani}, C. and {Pagano}, I. and {Pailler}, F. and
         {Palacin}, H. and {Palaversa}, L. and {Parsons}, P. and {Paulsen}, T. and
         {Pecoraro}, M. and {Pedrosa}, R. and {Pentik{\"a}inen}, H. and
         {Pereira}, J. and {Pichon}, B. and {Piersimoni}, A.~M. and
         {Pineau}, F. -X. and {Plachy}, E. and {Plum}, G. and {Poujoulet}, E. and
         {Pr{\v{s}}a}, A. and {Pulone}, L. and {Ragaini}, S. and {Rago}, S. and
         {Rambaux}, N. and {Ramos-Lerate}, M. and {Ranalli}, P. and {Rauw}, G. and
         {Read}, A. and {Regibo}, S. and {Renk}, F. and {Reyl{\'e}}, C. and
         {Ribeiro}, R.~A. and {Rimoldini}, L. and {Ripepi}, V. and {Riva}, A. and
         {Rixon}, G. and {Roelens}, M. and {Romero-G{\'o}mez}, M. and
         {Rowell}, N. and {Royer}, F. and {Rudolph}, A. and {Ruiz-Dern}, L. and
         {Sadowski}, G. and {Sagrist{\`a} Sell{\'e}s}, T. and {Sahlmann}, J. and
         {Salgado}, J. and {Salguero}, E. and {Sarasso}, M. and {Savietto}, H. and
         {Schnorhk}, A. and {Schultheis}, M. and {Sciacca}, E. and {Segol}, M. and
         {Segovia}, J.~C. and {Segransan}, D. and {Serpell}, E. and
         {Shih}, I. -C. and {Smareglia}, R. and {Smart}, R.~L. and {Smith}, C. and
         {Solano}, E. and {Solitro}, F. and {Sordo}, R. and {Soria Nieto}, S. and
         {Souchay}, J. and {Spagna}, A. and {Spoto}, F. and {Stampa}, U. and
         {Steele}, I.~A. and {Steidelm{\"u}ller}, H. and {Stephenson}, C.~A. and
         {Stoev}, H. and {Suess}, F.~F. and {S{\"u}veges}, M. and {Surdej}, J. and
         {Szabados}, L. and {Szegedi-Elek}, E. and {Tapiador}, D. and
         {Taris}, F. and {Tauran}, G. and {Taylor}, M.~B. and {Teixeira}, R. and
         {Terrett}, D. and {Tingley}, B. and {Trager}, S.~C. and {Turon}, C. and
         {Ulla}, A. and {Utrilla}, E. and {Valentini}, G. and {van Elteren}, A. and
         {Van Hemelryck}, E. and {van Leeuwen}, M. and {Varadi}, M. and
         {Vecchiato}, A. and {Veljanoski}, J. and {Via}, T. and {Vicente}, D. and
         {Vogt}, S. and {Voss}, H. and {Votruba}, V. and {Voutsinas}, S. and
         {Walmsley}, G. and {Weiler}, M. and {Weingrill}, K. and {Werner}, D. and
         {Wevers}, T. and {Whitehead}, G. and {Wyrzykowski}, {\L}. and
         {Yoldas}, A. and {{\v{Z}}erjal}, M. and {Zucker}, S. and {Zurbach}, C. and
         {Zwitter}, T. and {Alecu}, A. and {Allen}, M. and {Allende Prieto}, C. and
         {Amorim}, A. and {Anglada-Escud{\'e}}, G. and {Arsenijevic}, V. and
         {Azaz}, S. and {Balm}, P. and {Beck}, M. and {Bernstein}, H. -H. and
         {Bigot}, L. and {Bijaoui}, A. and {Blasco}, C. and {Bonfigli}, M. and
         {Bono}, G. and {Boudreault}, S. and {Bressan}, A. and {Brown}, S. and
         {Brunet}, P. -M. and {Bunclark}, P. and {Buonanno}, R. and
         {Butkevich}, A.~G. and {Carret}, C. and {Carrion}, C. and {Chemin}, L. and
         {Ch{\'e}reau}, F. and {Corcione}, L. and {Darmigny}, E. and
         {de Boer}, K.~S. and {de Teodoro}, P. and {de Zeeuw}, P.~T. and
         {Delle Luche}, C. and {Domingues}, C.~D. and {Dubath}, P. and
         {Fodor}, F. and {Fr{\'e}zouls}, B. and {Fries}, A. and {Fustes}, D. and
         {Fyfe}, D. and {Gallardo}, E. and {Gallegos}, J. and {Gardiol}, D. and
         {Gebran}, M. and {Gomboc}, A. and {G{\'o}mez}, A. and {Grux}, E. and
         {Gueguen}, A. and {Heyrovsky}, A. and {Hoar}, J. and {Iannicola}, G. and
         {Isasi Parache}, Y. and {Janotto}, A. -M. and {Joliet}, E. and
         {Jonckheere}, A. and {Keil}, R. and {Kim}, D. -W. and {Klagyivik}, P. and
         {Klar}, J. and {Knude}, J. and {Kochukhov}, O. and {Kolka}, I. and
         {Kos}, J. and {Kutka}, A. and {Lainey}, V. and {LeBouquin}, D. and
         {Liu}, C. and {Loreggia}, D. and {Makarov}, V.~V. and
         {Marseille}, M.~G. and {Martayan}, C. and {Martinez-Rubi}, O. and
         {Massart}, B. and {Meynadier}, F. and {Mignot}, S. and {Munari}, U. and
         {Nguyen}, A. -T. and {Nordlander}, T. and {Ocvirk}, P. and
         {O'Flaherty}, K.~S. and {Olias Sanz}, A. and {Ortiz}, P. and
         {Osorio}, J. and {Oszkiewicz}, D. and {Ouzounis}, A. and {Palmer}, M. and
         {Park}, P. and {Pasquato}, E. and {Peltzer}, C. and {Peralta}, J. and
         {P{\'e}turaud}, F. and {Pieniluoma}, T. and {Pigozzi}, E. and
         {Poels}, J. and {Prat}, G. and {Prod'homme}, T. and {Raison}, F. and
         {Rebordao}, J.~M. and {Risquez}, D. and {Rocca-Volmerange}, B. and
         {Rosen}, S. and {Ruiz-Fuertes}, M.~I. and {Russo}, F. and {Sembay}, S. and
         {Serraller Vizcaino}, I. and {Short}, A. and {Siebert}, A. and
         {Silva}, H. and {Sinachopoulos}, D. and {Slezak}, E. and {Soffel}, M. and
         {Sosnowska}, D. and {Strai{\v{z}}ys}, V. and {ter Linden}, M. and
         {Terrell}, D. and {Theil}, S. and {Tiede}, C. and {Troisi}, L. and
         {Tsalmantza}, P. and {Tur}, D. and {Vaccari}, M. and {Vachier}, F. and
         {Valles}, P. and {Van Hamme}, W. and {Veltz}, L. and {Virtanen}, J. and
         {Wallut}, J. -M. and {Wichmann}, R. and {Wilkinson}, M.~I. and
         {Ziaeepour}, H. and {Zschocke}, S.},
        title = "{The Gaia mission}",
      journal = {\aap},
         year = "2016",
        month = "Nov",
       volume = {595},
          eid = {A1},
        pages = {A1},
          doi = {10.1051/0004-6361/201629272},
archivePrefix = {arXiv},
       eprint = {1609.04153},
 primaryClass = {astro-ph.IM},
       adsurl = {https://ui.adsabs.harvard.edu/abs/2016A&A...595A...1G}
}

@ARTICLE{gai18,
   author = {{Gaia Collaboration} and {Brown}, A.~G.~A. and {Vallenari}, A. and
	{Prusti}, T. and {de Bruijne}, J.~H.~J. and {Babusiaux}, C. and
	{Bailer-Jones}, C.~A.~L. and {Biermann}, M. and {Evans}, D.~W. and
	{Eyer}, L. and et al.},
    title = "{Gaia Data Release 2. Summary of the contents and survey properties}",
  journal = {\aap},
archivePrefix = "arXiv",
   eprint = {1804.09365},
     year = 2018,
    month = aug,
   volume = 616,
      eid = {A1},
    pages = {A1},
      doi = {10.1051/0004-6361/201833051},
   adsurl = {http://adsabs.harvard.edu/abs/2018A%26A...616A...1G}
}

@ARTICLE{jer23,
       author = {{Jermyn}, Adam S. and {Bauer}, Evan B. and {Schwab}, Josiah and {Farmer}, R. and {Ball}, Warrick H. and {Bellinger}, Earl P. and {Dotter}, Aaron and {Joyce}, Meridith and {Marchant}, Pablo and {Mombarg}, Joey S.~G. and {Wolf}, William M. and {Sunny Wong}, Tin Long and {Cinquegrana}, Giulia C. and {Farrell}, Eoin and {Smolec}, R. and {Thoul}, Anne and {Cantiello}, Matteo and {Herwig}, Falk and {Toloza}, Odette and {Bildsten}, Lars and {Townsend}, Richard H.~D. and {Timmes}, F.~X.},
        title = "{Modules for Experiments in Stellar Astrophysics (MESA): Time-dependent Convection, Energy Conservation, Automatic Differentiation, and Infrastructure}",
      journal = {\apjs},
         year = 2023,
        month = mar,
       volume = {265},
       number = {1},
          eid = {15},
        pages = {15},
          doi = {10.3847/1538-4365/acae8d},
archivePrefix = {arXiv},
       eprint = {2208.03651},
 primaryClass = {astro-ph.SR},
       adsurl = {https://ui.adsabs.harvard.edu/abs/2023ApJS..265...15J}
}

@MISC{mor15,
   author = {{Morton}, T.~D.},
    title = "{isochrones: Stellar model grid package}",
howpublished = {Astrophysics Source Code Library},
     year = 2015,
    month = mar,
      url = {https://isochrones.readthedocs.io/en/latest/},
archivePrefix = "ascl",
   eprint = {1503.010},
   adsurl = {https://ui.adsabs.harvard.edu/abs/2015ascl.soft03010M}
}

@ARTICLE{pax11,
       author = {{Paxton}, Bill and {Bildsten}, Lars and {Dotter}, Aaron and
         {Herwig}, Falk and {Lesaffre}, Pierre and {Timmes}, Frank},
        title = "{Modules for Experiments in Stellar Astrophysics (MESA)}",
      journal = {\apjs},
         year = 2011,
        month = jan,
       volume = {192},
       number = {1},
          eid = {3},
        pages = {3},
          doi = {10.1088/0067-0049/192/1/3},
archivePrefix = {arXiv},
       eprint = {1009.1622},
 primaryClass = {astro-ph.SR},
       adsurl = {https://ui.adsabs.harvard.edu/abs/2011ApJS..192....3P}
}

@ARTICLE{pax13,
       author = {{Paxton}, Bill and {Cantiello}, Matteo and {Arras}, Phil and
         {Bildsten}, Lars and {Brown}, Edward F. and {Dotter}, Aaron and
         {Mankovich}, Christopher and {Montgomery}, M.~H. and {Stello}, Dennis and
         {Timmes}, F.~X. and {Townsend}, Richard},
        title = "{Modules for Experiments in Stellar Astrophysics (MESA): Planets, Oscillations, Rotation, and Massive Stars}",
      journal = {\apjs},
         year = 2013,
        month = sep,
       volume = {208},
       number = {1},
          eid = {4},
        pages = {4},
          doi = {10.1088/0067-0049/208/1/4},
archivePrefix = {arXiv},
       eprint = {1301.0319},
 primaryClass = {astro-ph.SR},
       adsurl = {https://ui.adsabs.harvard.edu/abs/2013ApJS..208....4P}
}

@ARTICLE{pax18,
       author = {{Paxton}, Bill and {Schwab}, Josiah and {Bauer}, Evan B. and
         {Bildsten}, Lars and {Blinnikov}, Sergei and {Duffell}, Paul and
         {Farmer}, R. and {Goldberg}, Jared A. and {Marchant}, Pablo and
         {Sorokina}, Elena and {Thoul}, Anne and {Townsend}, Richard H.~D. and
         {Timmes}, F.~X.},
        title = "{Modules for Experiments in Stellar Astrophysics (MESA): Convective Boundaries, Element Diffusion, and Massive Star Explosions}",
      journal = {\apjs},
         year = 2018,
        month = feb,
       volume = {234},
       number = {2},
          eid = {34},
        pages = {34},
          doi = {10.3847/1538-4365/aaa5a8},
archivePrefix = {arXiv},
       eprint = {1710.08424},
 primaryClass = {astro-ph.SR},
       adsurl = {https://ui.adsabs.harvard.edu/abs/2018ApJS..234...34P}
}

@ARTICLE{pax19,
       author = {{Paxton}, Bill and {Smolec}, R. and {Schwab}, Josiah and {Gautschy}, A. and {Bildsten}, Lars and {Cantiello}, Matteo and {Dotter}, Aaron and {Farmer}, R. and {Goldberg}, Jared A. and {Jermyn}, Adam S. and {Kanbur}, S.~M. and {Marchant}, Pablo and {Thoul}, Anne and {Townsend}, Richard H.~D. and {Wolf}, William M. and {Zhang}, Michael and {Timmes}, F.~X.},
        title = "{Modules for Experiments in Stellar Astrophysics (MESA): Pulsating Variable Stars, Rotation, Convective Boundaries, and Energy Conservation}",
      journal = {\apjs},
         year = 2019,
        month = jul,
       volume = {243},
       number = {1},
          eid = {10},
        pages = {10},
          doi = {10.3847/1538-4365/ab2241},
archivePrefix = {arXiv},
       eprint = {1903.01426},
 primaryClass = {astro-ph.SR},
       adsurl = {https://ui.adsabs.harvard.edu/abs/2019ApJS..243...10P}
}

@ARTICLE{Rustamkulov2022ApJ...928L...7R,
       author = {{Rustamkulov}, Zafar and {Sing}, David K. and {Liu}, Rongrong and {Wang}, Ashley},
        title = "{Analysis of a JWST NIRSpec Lab Time Series: Characterizing Systematics, Recovering Exoplanet Transit Spectroscopy, and Constraining a Noise Floor}",
      journal = {\apjl},
         year = 2022,
        month = mar,
       volume = {928},
       number = {1},
          eid = {L7},
        pages = {L7},
          doi = {10.3847/2041-8213/ac5b6f},
archivePrefix = {arXiv},
       eprint = {2203.04173},
 primaryClass = {astro-ph.EP},
       adsurl = {https://ui.adsabs.harvard.edu/abs/2022ApJ...928L...7R}
}

@ARTICLE{Rustamkulov2023Natur.614..659R,
       author = {{Rustamkulov}, Z. and {Sing}, D.~K. and {Mukherjee}, S. and {May}, E.~M. and {Kirk}, J. and {Schlawin}, E. and {Line}, M.~R. and {Piaulet}, C. and {Carter}, A.~L. and {Batalha}, N.~E. and {Goyal}, J.~M. and {L{\'o}pez-Morales}, M. and {Lothringer}, J.~D. and {MacDonald}, R.~J. and {Moran}, S.~E. and {Stevenson}, K.~B. and {Wakeford}, H.~R. and {Espinoza}, N. and {Bean}, J.~L. and {Batalha}, N.~M. and {Benneke}, B. and {Berta-Thompson}, Z.~K. and {Crossfield}, I.~J.~M. and {Gao}, P. and {Kreidberg}, L. and {Powell}, D.~K. and {Cubillos}, P.~E. and {Gibson}, N.~P. and {Leconte}, J. and {Molaverdikhani}, K. and {Nikolov}, N.~K. and {Parmentier}, V. and {Roy}, P. and {Taylor}, J. and {Turner}, J.~D. and {Wheatley}, P.~J. and {Aggarwal}, K. and {Ahrer}, E. and {Alam}, M.~K. and {Alderson}, L. and {Allen}, N.~H. and {Banerjee}, A. and {Barat}, S. and {Barrado}, D. and {Barstow}, J.~K. and {Bell}, T.~J. and {Blecic}, J. and {Brande}, J. and {Casewell}, S. and {Changeat}, Q. and {Chubb}, K.~L. and {Crouzet}, N. and {Daylan}, T. and {Decin}, L. and {D{\'e}sert}, J. and {Mikal-Evans}, T. and {Feinstein}, A.~D. and {Flagg}, L. and {Fortney}, J.~J. and {Harrington}, J. and {Heng}, K. and {Hong}, Y. and {Hu}, R. and {Iro}, N. and {Kataria}, T. and {Kempton}, E.~M. -R. and {Krick}, J. and {Lendl}, M. and {Lillo-Box}, J. and {Louca}, A. and {Lustig-Yaeger}, J. and {Mancini}, L. and {Mansfield}, M. and {Mayne}, N.~J. and {Miguel}, Y. and {Morello}, G. and {Ohno}, K. and {Palle}, E. and {Petit dit de la Roche}, D.~J.~M. and {Rackham}, B.~V. and {Radica}, M. and {Ramos-Rosado}, L. and {Redfield}, S. and {Rogers}, L.~K. and {Shkolnik}, E.~L. and {Southworth}, J. and {Teske}, J. and {Tremblin}, P. and {Tucker}, G.~S. and {Venot}, O. and {Waalkes}, W.~C. and {Welbanks}, L. and {Zhang}, X. and {Zieba}, S.},
        title = "{Early Release Science of the exoplanet WASP-39b with JWST NIRSpec PRISM}",
      journal = {\nat},
         year = 2023,
        month = feb,
       volume = {614},
       number = {7949},
        pages = {659-663},
          doi = {10.1038/s41586-022-05677-y},
archivePrefix = {arXiv},
       eprint = {2211.10487},
 primaryClass = {astro-ph.EP},
       adsurl = {https://ui.adsabs.harvard.edu/abs/2023Natur.614..659R}
}

@ARTICLE{2025ApJ...993...78S,
       author = {{Sodickson}, Noah and {Grunblatt}, Samuel},
        title = "{In Search of Decay: An Analysis of Transit Times of Hot Jupiters in Main-sequence and Post-main-sequence Systems}",
      journal = {\apj},
         year = 2025,
        month = nov,
       volume = {993},
       number = {1},
          eid = {78},
        pages = {78},
          doi = {10.3847/1538-4357/adfd60},
archivePrefix = {arXiv},
       eprint = {2508.18355},
 primaryClass = {astro-ph.EP},
       adsurl = {https://ui.adsabs.harvard.edu/abs/2025ApJ...993...78S}
}

@ARTICLE{2022ApJS..259...62I,
       author = {{Ivshina}, Ekaterina S. and {Winn}, Joshua N.},
        title = "{TESS Transit Timing of Hundreds of Hot Jupiters}",
      journal = {\apjs},
         year = 2022,
        month = apr,
       volume = {259},
       number = {2},
          eid = {62},
        pages = {62},
          doi = {10.3847/1538-4365/ac545b},
archivePrefix = {arXiv},
       eprint = {2202.03401},
 primaryClass = {astro-ph.EP},
       adsurl = {https://ui.adsabs.harvard.edu/abs/2022ApJS..259...62I}
}

@ARTICLE{2023ApJS..265....4K,
       author = {{Kokori}, A. and {Tsiaras}, A. and {Edwards}, B. and {Jones}, A. and {Pantelidou}, G. and {Tinetti}, G. and {Bewersdorff}, L. and {Iliadou}, A. and {Jongen}, Y. and {Lekkas}, G. and {Nastasi}, A. and {Poultourtzidis}, E. and {Sidiropoulos}, C. and {Walter}, F. and {W{\"u}nsche}, A. and {Abraham}, R. and {Agnihotri}, V.~K. and {Albanesi}, R. and {Arce-Mansego}, E. and {Arnot}, D. and {Audejean}, M. and {Aumasson}, C. and {Bachschmidt}, M. and {Baj}, G. and {Barroy}, P.~R. and {Belinski}, A.~A. and {Bennett}, D. and {Benni}, P. and {Bernacki}, K. and {Betti}, L. and {Biagini}, A. and {Bosch}, P. and {Brandebourg}, P. and {Br{\'a}t}, L. and {Bretton}, M. and {Brincat}, S.~M. and {Brouillard}, S. and {Bruzas}, A. and {Bruzzone}, A. and {Buckland}, R.~A. and {Cal{\'o}}, M. and {Campos}, F. and {Carre{\~n}o}, A. and {Carrion Rodrigo}, J.~A. and {Casali}, R. and {Casalnuovo}, G. and {Cataneo}, M. and {Chang}, C.-M. and {Changeat}, L. and {Chowdhury}, V. and {Ciantini}, R. and {Cilluffo}, M. and {Coliac}, J.-F. and {Conzo}, G. and {Correa}, M. and {Coulon}, G. and {Crouzet}, N. and {Crow}, M.~V. and {Curtis}, I.~A. and {Daniel}, D. and {Dauchet}, B. and {Dawes}, S. and {Deldem}, M. and {Deligeorgopoulos}, D. and {Dransfield}, G. and {Dymock}, R. and {Eenm{\"a}e}, T. and {Esseiva}, N. and {Evans}, P. and {Falco}, C. and {Farf{\'a}n}, R.~G. and {Fern{\'a}ndez-Laj{\'u}s}, E. and {Ferratfiat}, S. and {Ferreira}, S.~L. and {Ferretti}, A. and {Fio{\l}ka}, J. and {Fowler}, M. and {Futcher}, S.~R. and {Gabellini}, D. and {Gainey}, T. and {Gaitan}, J. and {Gajdo{\v{s}}}, P. and {Garc{\'\i}a-S{\'a}nchez}, A. and {Garlitz}, J. and {Gillier}, C. and {Gison}, C. and {Gonzales}, J. and {Gorshanov}, D. and {Grau Horta}, F. and {Grivas}, G. and {Guerra}, P. and {Guillot}, T. and {Haswell}, C.~A. and {Haymes}, T. and {Hentunen}, V.-P. and {Hills}, K. and {Hose}, K. and {Humbert}, T. and {Hurter}, F. and {Hynek}, T. and {Irzyk}, M. and {Jacobsen}, J. and {Jannetta}, A.~L. and {Johnson}, K. and {J{\'o}{\'z}wik-Wabik}, P. and {Kaeouach}, A.~E. and {Kang}, W. and {Kiiskinen}, H. and {Kim}, T. and {Kivila}, {\"U}. and {Koch}, B. and {Kolb}, U. and {Ku{\v{c}}{\'a}kov{\'a}}, H. and {Lai}, S.-P. and {Laloum}, D. and {Lasota}, S. and {Lewis}, L.~A. and {Liakos}, G.-I. and {Libotte}, F. and {Lomoz}, F. and {Lopresti}, C. and {Majewski}, R. and {Malcher}, A. and {Mallonn}, M. and {Mannucci}, M. and {Marchini}, A. and {Mari}, J.-M. and {Marino}, A. and {Marino}, G. and {Mario}, J.-C. and {Marquette}, J.-B. and {Mart{\'\i}nez-Bravo}, F.~A. and {Ma{\v{s}}ek}, M. and {Matassa}, P. and {Michel}, P. and {Michelet}, J. and {Miller}, M. and {Miny}, E. and {Molina}, D. and {Mollier}, T. and {Monteleone}, B. and {Montigiani}, N. and {Morales-Aimar}, M. and {Mortari}, F. and {Morvan}, M. and {Mugnai}, L.~V. and {Murawski}, G. and {Naponiello}, L. and {Naudin}, J.-L. and {Naves}, R. and {N{\'e}el}, D. and {Neito}, R. and {Neveu}, S. and {Noschese}, A. and {{\"O}{\u{g}}men}, Y. and {Ohshima}, O. and {Orbanic}, Z. and {Pace}, E.~P. and {Pantacchini}, C. and {Paschalis}, N.~I. and {Pereira}, C. and {Peretto}, I. and {Perroud}, V. and {Phillips}, M. and {Pintr}, P. and {Pioppa}, J.-B. and {Plazas}, J. and {Poelarends}, A.~J. and {Popowicz}, A. and {Purcell}, J. and {Quinn}, N. and {Raetz}, M. and {Rees}, D. and {Regembal}, F. and {Rocchetto}, M. and {Rocci}, P.-F. and {Rockenbauer}, M. and {Roth}, R. and {Rousselot}, L. and {Rubia}, X. and {Ruocco}, N. and {Russo}, E. and {Salisbury}, M. and {Salvaggio}, F. and {Santos}, A. and {Savage}, J. and {Scaggiante}, F. and {Sedita}, D. and {Shadick}, S. and {Silva}, A.~F. and {Sioulas}, N. and {{\v{S}}koln{\'\i}k}, V. and {Smith}, M. and {Smolka}, M. and {Solmaz}, A. and {Stanbury}, N. and {Stouraitis}, D. and {Tan}, T.-G. and {Theusner}, M. and {Thurston}, G.},
        title = "{ExoClock Project. III. 450 New Exoplanet Ephemerides from Ground and Space Observations}",
      journal = {\apjs},
         year = 2023,
        month = mar,
       volume = {265},
       number = {1},
          eid = {4},
        pages = {4},
          doi = {10.3847/1538-4365/ac9da4},
archivePrefix = {arXiv},
       eprint = {2209.09673},
 primaryClass = {astro-ph.EP},
       adsurl = {https://ui.adsabs.harvard.edu/abs/2023ApJS..265....4K}
}

@ARTICLE{Borsa2021,
       author = {{Borsa}, F. and {Allart}, R. and {Casasayas-Barris}, N. and {Tabernero}, H. and {Zapatero Osorio}, M.~R. and {Cristiani}, S. and {Pepe}, F. and {Rebolo}, R. and {Santos}, N.~C. and {Adibekyan}, V. and {Bourrier}, V. and {Demangeon}, O.~D.~S. and {Ehrenreich}, D. and {Pall{\'e}}, E. and {Sousa}, S. and {Lillo-Box}, J. and {Lovis}, C. and {Micela}, G. and {Oshagh}, M. and {Poretti}, E. and {Sozzetti}, A. and {Allende Prieto}, C. and {Alibert}, Y. and {Amate}, M. and {Benz}, W. and {Bouchy}, F. and {Cabral}, A. and {Dekker}, H. and {D'Odorico}, V. and {Di Marcantonio}, P. and {Figueira}, P. and {Genova Santos}, R. and {Gonz{\'a}lez Hern{\'a}ndez}, J.~I. and {Lo Curto}, G. and {Manescau}, A. and {Martins}, C.~J.~A.~P. and {M{\'e}gevand}, D. and {Mehner}, A. and {Molaro}, P. and {Nunes}, N.~J. and {Riva}, M. and {Su{\'a}rez Mascare{\~n}o}, A. and {Udry}, S. and {Zerbi}, F.},
        title = "{Atmospheric Rossiter-McLaughlin effect and transmission spectroscopy of WASP-121b with ESPRESSO}",
      journal = {\aap},
         year = 2021,
        month = jan,
       volume = {645},
          eid = {A24},
        pages = {A24},
          doi = {10.1051/0004-6361/202039344},
archivePrefix = {arXiv},
       eprint = {2011.01245},
 primaryClass = {astro-ph.EP},
       adsurl = {https://ui.adsabs.harvard.edu/abs/2021A&A...645A..24B}
}

@ARTICLE{2013A&A...553A...6H,
   author = {{Husser}, T.-O. and {Wende-von Berg}, S. and {Dreizler}, S. and 
	{Homeier}, D. and {Reiners}, A. and {Barman}, T. and {Hauschildt}, P.~H.
	},
    title = "{A new extensive library of PHOENIX stellar atmospheres and synthetic spectra}",
  journal = {\aap},
archivePrefix = "arXiv",
   eprint = {1303.5632},
 primaryClass = "astro-ph.SR",
     year = 2013,
    month = may,
   volume = 553,
      eid = {A6},
    pages = {A6},
      doi = {10.1051/0004-6361/201219058},
   adsurl = {http://adsabs.harvard.edu/abs/2013A%26A...553A...6H}
}

@ARTICLE{
MandelAgol2002,
   author = {{Mandel}, K. and {Agol}, E.},
    title = "{Analytic Light Curves for Planetary Transit Searches}",
  journal = {\apjl},
   eprint = {astro-ph/0210099},
     year = 2002,
    month = dec,
   volume = 580,
    pages = {L171-L175},
      doi = {10.1086/345520},
   adsurl = {http://adsabs.harvard.edu/abs/2002ApJ...580L.171M}
}

@ARTICLE{
Claret2000,
   author = {{Claret}, A.},
    title = "{A new non-linear limb-darkening law for LTE stellar atmosphere models. Calculations for -5.0 {\lt}= log[M/H] {\lt}= +1, 2000 K {\lt}= T$_{eff}$ {\lt}= 50000 K at several surface gravities}",
  journal = {\aap},
     year = 2000,
    month = nov,
   volume = 363,
    pages = {1081-1190},
   adsurl = {http://adsabs.harvard.edu/abs/2000A%26A...363.1081C}
}

@ARTICLE{
Charbonneau2000,
   author = {{Charbonneau}, D. and {Brown}, T.~M. and {Latham}, D.~W. and 
	{Mayor}, M.},
    title = "{Detection of Planetary Transits Across a Sun-like Star}",
  journal = {\apjl},
   eprint = {astro-ph/9911436},
     year = 2000,
    month = jan,
   volume = 529,
    pages = {L45-L48},
      doi = {10.1086/312457},
   adsurl = {http://adsabs.harvard.edu/abs/2000ApJ...529L..45C}
}

@ARTICLE{
Charbonneau2002,
   author = {{Charbonneau}, D. and {Brown}, T.~M. and {Noyes}, R.~W. and 
	{Gilliland}, R.~L.},
    title = "{Detection of an Extrasolar Planet Atmosphere}",
  journal = {\apj},
   eprint = {astro-ph/0111544},
     year = 2002,
    month = mar,
   volume = 568,
    pages = {377-384},
      doi = {10.1086/338770},
   adsurl = {http://adsabs.harvard.edu/abs/2002ApJ...568..377C}
}

@ARTICLE{
Knutson2007,
   author = {{Knutson}, H.~A. and {Charbonneau}, D. and {Noyes}, R.~W. and 
	{Brown}, T.~M. and {Gilliland}, R.~L.},
    title = "{Using Stellar Limb-Darkening to Refine the Properties of HD 209458b}",
  journal = {\apj},
   eprint = {astro-ph/0603542},
     year = 2007,
    month = jan,
   volume = 655,
    pages = {564-575},
      doi = {10.1086/510111},
   adsurl = {http://adsabs.harvard.edu/abs/2007ApJ...655..564K}
}

@ARTICLE{
Brown2001,
   author = {{Brown}, T.~M.},
    title = "{Transmission Spectra as Diagnostics of Extrasolar Giant Planet Atmospheres}",
  journal = {\apj},
   eprint = {astro-ph/0101307},
     year = 2001,
    month = jun,
   volume = 553,
    pages = {1006-1026},
      doi = {10.1086/320950},
   adsurl = {http://adsabs.harvard.edu/abs/2001ApJ...553.1006B}
}

@ARTICLE{
Sing2009,
   author = {{Sing}, D.~K. and {D{\'e}sert}, J.-M. and {Lecavelier Des Etangs}, A. and 
	{Ballester}, G.~E. and {Vidal-Madjar}, A. and {Parmentier}, V. and 
	{Hebrard}, G. and {Henry}, G.~W.},
    title = "{Transit spectrophotometry of the exoplanet HD 189733b. I. Searching for water but finding haze with HST NICMOS}",
  journal = {\aap},
archivePrefix = "arXiv",
   eprint = {0907.4991},
 primaryClass = "astro-ph.EP",
     year = 2009,
    month = oct,
   volume = 505,
    pages = {891-899},
      doi = {10.1051/0004-6361/200912776},
   adsurl = {http://adsabs.harvard.edu/abs/2009A%26A...505..891S}
}

@ARTICLE{2003ApJ...585.1038S,
       author = {{Seager}, S. and {Mall{\'e}n-Ornelas}, G.},
        title = "{A Unique Solution of Planet and Star Parameters from an Extrasolar Planet Transit Light Curve}",
      journal = {\apj},
         year = "2003",
        month = "Mar",
       volume = {585},
       number = {2},
        pages = {1038-1055},
          doi = {10.1086/346105},
archivePrefix = {arXiv},
       eprint = {astro-ph/0206228},
 primaryClass = {astro-ph},
       adsurl = {https://ui.adsabs.harvard.edu/abs/2003ApJ...585.1038S}
}

@Misc{numpy,
  author =    {Travis Oliphant},
  title =     {{NumPy}: A guide to {NumPy}},
  year =      {2006--},
  howpublished = {USA: Trelgol Publishing},
  url = "http://www.numpy.org/",
  note = {[Online; accessed <today>]}
 }

@MISC{scipy,
       author = {{Virtanen}, Pauli and {Gommers}, Ralf and {Burovski}, Evgeni and
         {Oliphant}, Travis E. and {Cournapeau}, David and {Weckesser}, Warren and
         {alexbrc} and {Peterson}, Pearu and {endolith} and {Mayorov}, Nikolay and
         {van der Walt}, Stefan and {Wilson}, Josh and {Laxalde}, Denis and
         {Brett}, Matthew and {Millman}, Jarrod and {Lars} and {Nelson}, Andrew and
         {Haberland}, Matt and {eric-jones} and {Polat}, Ilhan and
         {Larson}, Eric and {Kern}, Robert and {Moore}, Eric and {Carey}, CJ and
         {Leslie}, Tim and {Perktold}, Josef and {Reddy}, Tyler and
         {Bharti}, Aditya and {Feng}, Yu and {Vanderplas}, Jake},
        title = "{scipy/scipy: SciPy 1.2.1}",
         year = "2019",
        month = "Feb",
          eid = {10.5281/zenodo.2560881},
          doi = {10.5281/zenodo.2560881},
      version = {v1.2.1},
    publisher = {Zenodo},
       adsurl = {https://ui.adsabs.harvard.edu/abs/2019zndo...2560881V}
}

@MISC{matplotlib,
       author = {{Caswell}, Thomas A and {Droettboom}, Michael and {Hunter}, John and
         {Firing}, Eric and {Lee}, Antony and {Klymak}, Jody and
         {Stansby}, David and {Sales de Andrade}, Elliott and
         {Hedegaard Nielsen}, Jens and {Varoquaux}, Nelle and {Root}, Benjamin and
         {Hoffmann}, Tim and {Elson}, Phil and {May}, Ryan and {Dale}, Darren and
         {Lee}, Jae-Joon and {Sepp{\"a}nen}, Jouni K. and {McDougall}, Damon and
         {Straw}, Andrew and {Hobson}, Paul and {Gohlke}, Christoph and
         {Yu}, Tony S and {Ma}, Eric and {Vincent}, Adrien F. and
         {Silvester}, Steven and {Moad}, Charlie and {Katins}, Jan and
         {Kniazev}, Nikita and {Ariza}, Federico and {Ernest}, Elan},
        title = "{matplotlib/matplotlib v3.1.0}",
         year = "2019",
        month = "May",
          eid = {10.5281/zenodo.2893252},
          doi = {10.5281/zenodo.2893252},
      version = {v3.1.0},
    publisher = {Zenodo},
       adsurl = {https://ui.adsabs.harvard.edu/abs/2019zndo...2893252C}
}

@ARTICLE{astropy,
       author = {{Astropy Collaboration} and {Price-Whelan}, A.~M. and
         {Sip{\H{o}}cz}, B.~M. and {G{\"u}nther}, H.~M. and {Lim}, P.~L. and
         {Crawford}, S.~M. and {Conseil}, S. and {Shupe}, D.~L. and
         {Craig}, M.~W. and {Dencheva}, N. and {Ginsburg}, A. and {Vand
        erPlas}, J.~T. and {Bradley}, L.~D. and {P{\'e}rez-Su{\'a}rez}, D. and
         {de Val-Borro}, M. and {Aldcroft}, T.~L. and {Cruz}, K.~L. and
         {Robitaille}, T.~P. and {Tollerud}, E.~J. and et al. and {Astropy Contributors}},
        title = "{The Astropy Project: Building an Open-science Project and Status of the v2.0 Core Package}",
      journal = {\aj},
         year = "2018",
        month = "Sep",
       volume = {156},
       number = {3},
          eid = {123},
        pages = {123},
          doi = {10.3847/1538-3881/aabc4f},
archivePrefix = {arXiv},
       eprint = {1801.02634},
 primaryClass = {astro-ph.IM},
       adsurl = {https://ui.adsabs.harvard.edu/abs/2018AJ....156..123A}
}

@article{sing2016_nature,
	Author = {Sing, David K. and Fortney, Jonathan J. and Nikolov, Nikolay and Wakeford, Hannah R. and Kataria, Tiffany and Evans, Thomas M. and Aigrain, Suzanne and Ballester, Gilda E. and Burrows, Adam S. and Deming, Drake and D{\'e}sert, Jean-Michel and Gibson, Neale P. and Henry, Gregory W. and Huitson, Catherine M. and Knutson, Heather A. and Etangs, Alain Lecavelier des and Pont, Frederic and Showman, Adam P. and Vidal-Madjar, Alfred and Williamson, Michael H. and Wilson, Paul A.},
	Date = {2016/01/07/print},
	Day = {07},
	Isbn = {0028-0836},
	Journal = {Nature},
	L3 = {10.1038/nature16068},
	M3 = {Letter},
	Month = {01},
	Number = {7584},
	Pages = {59--62},
	Publisher = {Nature Publishing Group, a division of Macmillan Publishers Limited. All Rights Reserved.},
	Title = {A continuum from clear to cloudy hot-Jupiter exoplanets without primordial water depletion},
	Ty = {JOUR},
	Url = {http://dx.doi.org/10.1038/nature16068},
	Volume = {529},
	Year = {2016},
}
\bibliographystyle{aasjournal}



\appendix
The complete machine-readable version of Table \ref{tab:mueff} provides the effective limb angle $\mu_{\rm eff}$ for each model in the PHOENIX spherical atmosphere grid of \citet{husser2013new}. The models span a $T_{\rm eff}$ from 2300–12000 K, $\log g$ from 0.0–6.0, and [M/H] from $-4.0$ to $+1.0$. These $\mu_{\rm eff}$ values, calculated using the $\tau_{\rm slant}=1$ method, enable rescaling of limb angles when computing limb-darkening coefficients from spherical models.

\begin{deluxetable*}{lcccc}
\tablecaption{Excerpt of the PHOENIX spherical atmosphere models and the corresponding
effective limb angles, $\mu_{\rm eff}$, computed using the physically motivated
$\tau_{\rm slant}=1$ rescaling method described in Section~\ref{sec:mueff}.
\label{tab:mueff}}
\tablehead{
\colhead{Filename} &
\colhead{$T_{\rm eff}$} &
\colhead{$\log g$} &
\colhead{[M/H]} &
\colhead{$\mu_{\rm eff}$}
}
\startdata
lte02300-0.00-0.0... & 2300 & 0.0 & 0.0 & 0.2934 \\
lte02300-0.50-0.0... & 2300 & 0.5 & 0.0 & 0.2346 \\
lte02300-1.00-0.0... & 2300 & 1.0 & 0.0 & 0.1915 \\
lte02300-1.50-0.0... & 2300 & 1.5 & 0.0 & 0.1458 \\
lte02300-2.00-0.0... & 2300 & 2.0 & 0.0 & 0.1293 \\
lte02300-2.50-0.0... & 2300 & 2.5 & 0.0 & 0.1054 \\
lte02300-3.00-0.0... & 2300 & 3.0 & 0.0 & 0.0783 \\
lte02300-3.50-0.0... & 2300 & 3.5 & 0.0 & 0.0592 \\
lte02300-4.00-0.0... & 2300 & 4.0 & 0.0 & 0.0436 \\
lte02300-4.50-0.0... & 2300 & 4.5 & 0.0 & 0.0342 \\
\label{tab:phoenix_mueff}
\enddata

\tablecomments{This table is published in its entirety in machine-readable format.
Columns list the PHOENIX atmosphere filename, effective temperature
($T_{\rm eff}$; K), surface gravity ($\log g$; cgs), metallicity ([M/H]; dex),
and the effective limb angle $\mu_{\rm eff}$ corresponding to the first
optically thick tangential ray ($\tau_{\rm slant}=1$). The full machine-readable
table contains $\mu_{\rm eff}$ values calculated for all successfully processed
models in the \citet{husser2013new} PHOENIX spherical atmosphere grid. A portion
is shown here for guidance regarding its form and content.}
\end{deluxetable*}

\end{document}